\input amstex
\input amsppt.sty

\hoffset=1.5cm

\TagsOnRight
\NoBlackBoxes

\define\Y{\Bbb Y}
\define\Z{\Bbb Z}
\define\C{\Bbb C}
\define\R{\Bbb R}
\define\al{\alpha}
\define\be{\beta}
\define\ga{\gamma}
\define\Ga{\Gamma}
\define\La{\Lambda}
\define\la{{\lambda}}

\define\th{{\theta}}
\define\om{\omega}
\define\Om{\Omega}
\define\LL{{\Cal L}}
\define\lv{l^{\varnothing}}

\define\bl{\bar l}
\define\bla{\overline\la}
\define\tE{E^{\#}_\th}
\define\wt{\widetilde}
\define\tht{\thetag}

\def\dimth{\varkappa_\theta}

\def\M{\underline{M}}
\def\wOm{\wt\Om}
\def\E{\Bbb E}
\def\dist{\operatorname{dist}}
\def\supp{\operatorname{supp}}
\def\const{\operatorname{const}}

\def\tth{\theta^{-1}}
\def\F{\Cal F}
\def\wh{\widehat}
\topmatter

\title Z-measures on partitions and their scaling limits
\endtitle
\author
Alexei Borodin and Grigori Olshanski
\endauthor
\date Preliminary version. October 25, 2002
\enddate

\abstract We study certain probability measures on partitions of $n=1,2,\dots$,
originated in representation theory, and demonstrate their connections with
random matrix theory and multivariate hypergeometric functions.

Our measures depend on three parameters including an analog of the $\be$
parameter in random matrix models. Under an appropriate limit transition as
$n\to\infty$, our measures converge to certain limit measures, which  are of
the same nature as one--dimensional log--gas with arbitrary $\be>0$.

The first main result says that averages of products of ``characteristic
polynomials'' with respect to the limit measures are given by the multivariate
hypergeometric functions of type (2,0). The second main result is a computation
of the limit correlation functions for the even values of $\be$.
\endabstract

\toc \widestnumber\head{6}
\head {} Introduction
\endhead
\head 1. Z--measures
\endhead
\head 2. Averages of $E_\th(\,\cdot\,;u_1)\cdots E_\th(\,\cdot\,;u_l)$ as
hypergeometric functions
\endhead
\head 3. Lattice correlation functions
\endhead
\head 4. Convergence of correlation functions
\endhead
\head 5. Limit correlation functions
\endhead
\head 6. Asymptotics of the correlation functions at the origin
\endhead
\head {} References \endhead
\endtoc

\endtopmatter

\document

\head Introduction
\endhead
The goal of this paper is to study certain measures on partitions which are in
many ways similar to log--gas (random matrix) models with arbitrary
$\beta=2\th$. The measures give rise to discrete (lattice) models. They admit
nontrivial scaling limits which have representation theoretic origin. The limit
objects can be viewed as random point processes on the real line.

In our earlier works \cite{P.I--P.V}, \cite{BO1--3}, \cite{Bor}, we thoroughly
studied the simplest case $\th=1$. In that case, the correlation functions in
the discrete and continuous pictures were explicitly computed in terms of the
Gauss hypergeometric function and the Whittaker function. Our goal is to see to
what extent these results can be carried over to the general $\th$.

As for the log--gas models, it seems to be very hard to compute the correlation
functions for general $\th$. However, one can evaluate other quantities of
interest. In \cite{Aom}, \cite{Ka}, \cite{BF} the authors computed the averages
of products of characteristic polynomials in random matrix type ensembles for
general $\th$. The answer is always given in terms of a multivariate
hypergeometric function.

Our first result is of the same kind: we show that in our model, the averaged
product of the natural analogs of characteristic polynomials is given by the
multivariate hypergeometric functions of type (2,1) or (2,0).

The main difference of our situation, as compared to random matrices,
is that we
are dealing with the infinite number of particles. In a degenerate situation,
our model turns into the Laguerre ensemble of the random matrix
theory, and we
recover known results of \cite{Ka}, \cite{BF}.

Our second result states that for integral $\th$ we can extract the correlation
functions of our measures from the averages of the ``characteristic
polynomials''. The correlation functions are given by hypergeometric functions
with repeated arguments. For similar results in the random matrix context, see
\cite{BF}, \cite{F1, section 4}, \cite{Ok1}, and references therein.

Finally, our third result is a computation of a scaling limit of the
correlation functions for integral $\th$. This limit transition is similar to
the bulk scaling limit in the random matrix ensembles. The limit correlation
functions are translation invariant and are given in terms of the $A$--type
spherical function of Heckman--Opdam.

The paper is organized as follows. In \S1 we introduce a family of measures on
partitions depending on two parameters and explain that these measures must have
a scaling limit as the size of partitions tends to infinity. In \S2 we compute,
in terms of hypergeometric functions, the averages of products of
``characteristic polynomials'' with respect to the limit measures. In \S3 we
relate, for the integral values of $\th$, the lattice correlation functions and
averages of analogs of characteristic polynomials for partitions. In \S4 we
prove that the lattice correlation functions converge to the correlation
functions of the limit measure in the appropriate scaling limit. In \S5 we
express the limit correlation functions through the hypergeometric functions. In
\S6 we compute the ``tail asymptotics'' of the limit correlation functions,
which leads to a translation invariant answer.

The authors are grateful to Peter Forrester for valuable remarks.

This research was partially conducted during the period the first author
(A.~B.) served as a Clay Mathematics Institute Long-Term Prize Fellow.

\head 1. Z--measures
\endhead

Let $\Y_n$ be the set of all partitions of a natural number $n$ (equivalently,
the set of all Young diagrams with $n$ boxes). For any $n=1,2,\dots$, we
consider a three-parameter family of probability measures $M^{(n)}_{z,z',\th}$
on $\Y_n$ given by
$$
M_{z,z',\th}^{(n)}(\la)
=\frac{n!\,(z)_{\lambda,\th}(z')_{\la,\th}}{(t)_nH(\la,\th)H'(\la,\th)}\,,
\tag 1.1
$$
where we use the following notation:

$z,z'\in\C$ and $\theta>0$ are parameters (admissible values of $(z,z')$ are
described below) and $t=zz'/\th$;

$\la$ is a Young diagram with $n$ boxes;

$$(t)_n=t(t+1)\cdots(t+n-1)=\frac{\Ga(t+n)}{\Gamma(t)}$$ is the Pochhammer
symbol;

$$
(z)_{\la,\theta}=\prod_{(i,j)\in\la}(z+(j-1)-(i-
1)\th)=\prod_{i=1}^{\ell(\la)}(z-(i-1)\th)_{\la_i}
$$
is a multidimensional analog of the Pochhammer symbol (here $(i,j)\in\lambda$
stands for the box in the $i$th row and $j$th column of the Young diagram $\la$,
and $\ell(\la)$ denotes the number of rows of $\la$);

$$
\gathered
H(\la,\th)=\prod_{(i,j)\in\la}((\la_i-j)+(\la_j'-i)\th+1),\\
H'(\la,\th)=\prod_{(i,j)\in\la}((\la_i-j)+(\la_j'-i)\th+\th),
\endgathered
$$
where $\la'$ denotes the transposed diagram.

One can easily see that
$$
M_{z,z',\th}^{(n)}(\la)=M_{-z/\th,-z'/\th,1/\th}^{(n)}(\la').
$$

Note that for any fixed $\la$, $M_{z,z',\th}^{(n)}(\la)$ is a rational function
in $z,z',\th$.

\proclaim{Proposition 1.1}
$$
\sum_{\la\in\Y_n}M_{z,z',\th}^{(n)}(\la)\equiv1.
$$
\endproclaim

\demo{Proof} See \cite{Ke2}, \cite{BO4}.
\enddemo

\proclaim{Proposition 1.2} The expression \tht{1.1} for
$M_{z,z',\th}^{(n)}(\la)$ is strictly positive for all $n=1,2,\dots$ and all
$\la\in\Y_n$ if and only if\/{\rm:}

$\bullet$ either $z\in\C\setminus (\Z_{\le0}+\Z_{\ge0}\,\th)$ and
$z'=\overline{z}$ {\rm(}the principal series{\rm)}

$\bullet$ or, under the additional assumption that $\th$ is rational, both
$z,z'$ are real numbers lying in one of the intervals between two consecutive
numbers from the lattice $\Z+\Z\th$ {\rm(}the complementary series{\rm)}.
\endproclaim

\demo{Proof} We have to find necessary and sufficient conditions under which
$$
\frac{\prod_{(i,j)\in\la}(z+c_\th(i,j))(z'+c_\th(i,j))}
{(zz')(zz'+\th)\dots(zz'+(n-1)\th)}>0, \qquad \text{where
$c_\th(i,j):=(j-1)-(i-1)\th$},
$$
for any $n=1,2,\dots$ and any $\la\in\Y_n$.  In the particular case $\th=1$
this was proved in \cite{P.I, Proposition 2.2}. The same argument works with
minor modifications.

Sufficiency: Our conditions imply that $(z+c_\th(i,j))(z'+c_\th(i,j))>0$ for
any $(i,j)$, so that the numerator is always strictly positive. They also imply
$zz'>0$, so that the denominator is strictly positive, too.

Necessity: For any $(i,j)$ and any $n$ large enough there exist diagrams
$\la\in\Y_n$ and $\mu\in\Y_{n-1}$ such that $\mu\subset\la$ and
$\la\setminus\mu=\{(i,j)\}$. Dividing the expression corresponding to $\la$ by
that corresponding to $\mu$ we see that
$$
\frac{(z+c)(z'+c)}{(zz'+n\th)}>0, \qquad c=c_\th(i,j).
$$
Note that $c$ can take any value from the set
$(\Z_{\ge0}+\Z_{\le0}\,\th)\subset\R$.

Letting $n\to\infty$ we conclude that the numerator $(z+c)(z'+c)$ must be real
and strictly positive for any $c$ from the set indicated above. It follows that
both $zz'$ and $z+z'$ are real. Hence, either $z,z'$ are complex conjugate to
each other or they are both real.

In the former case, the inequality $z+c\ne0$ implies that
$z\notin(\Z_{\le0}+\Z_{\ge0}\,\th)$. Hence, $z,z'$ are in the principal series.

In the latter case, we may assume that $z\ne z'$ (otherwise $z,z'$ are in the
principal series). We use the fact that $z+c$ and $z'+c$ must be of the same
sign for any $c$. If $\th$ is irrational then the numbers $c$ form an
everywhere dense subset in $\R$, so that there exists $c$ such that $-c$ is
strictly between $z$ and $z'$, which leads to a contradiction. Thus, $\th$ is
rational. Then $\Z_{\ge0}+\Z_{\le0}\,\th$ coincides with the lattice
$\Z+\Z\th$. Since $z,z'$ cannot be separated by a point of this lattice, we
conclude that $(z,z')$ is in the complementary series. \qed
\enddemo

In addition to the principal and complementary series of couples $(z,z')$ there
also exist $(z,z')$ such that the expression \tht{1.1} vanishes on a nonempty
subset of diagrams $\la$ and is strictly positive on the remaining diagrams. By
definition, such couples $(z,z')$ form the {\it degenerate series.\/} In the
next two propositions we provide examples of $(z,z')$ belonging to the
degenerate series.

\proclaim{Proposition 1.3} Let $m=1,2,\dots$, and assume that $z,z'$ satisfy
one of the following two conditions\/ {\rm (1), (2):}

{\rm(1)} {\rm(}$z=m\theta$, $z'>(m-1)\th${\rm)} {\rm or} {\rm(}$z'=m\theta$,
$z>(m-1)\th${\rm);}

{\rm(2)} {\rm(}$z=-m$, $z'<-m+1${\rm)} {\rm or} {\rm(}$z'=-m$, $z<-m+1${\rm)}.

Then $(z,z')$ is in the degenerate series. The set of diagrams $\la$ such that
the expression \tht{1.1} is strictly positive looks, respectively, as
follows{\rm:}

{\rm(1)} all diagrams with at most $m$ rows\/{\rm;}

{\rm(2)} all diagrams with at most $m$ columns.

\endproclaim

\demo{Proof} We leave the proof to the reader. \qed
\enddemo

Given $k,l\in\{1,2,\dots\}$, let $\Ga(k,l)$ denote the set of all boxes $(i,j)$
such that at least one of the inequalities $i\le k$, $j\le l$ holds (a ``fat
hook shape'').

\proclaim{Proposition 1.4} If $\th$ is irrational, let $k,l\in\{1,2,\dots\}$ be
arbitrary. If $\th$ is a rational number not equal to 1, write it as the ratio
$\th=s/r$ of relatively prime natural numbers, and then assume that at least
one of the inequalities $k<r$, $l<s$ holds. Finally, if $\th=1$ then assume
$k=l=1$.

Under these assumptions, assume further that both parameters $z,z'$ are real,
one of them equals $-(k-l\th)$, and the difference $|z-z'|$ is small enough.

Then $(z,z')$ is in the degenerate series, and the expression \tht{1.1} is
strictly positive exactly on whose diagrams that are contained in the ``fat
hook shape'' $\Ga(k,l)$ as defined above.
\endproclaim

\demo{Proof} We leave the proof to the reader. \qed
\enddemo

Thus, if the parameters $z,z'$ are in the principal, complementary, or
degenerate series then $M_{z,z',\th}^{(n)}$ is a probability measure on $\Y_n$
for any $n=1,2,\dots$\,. These measures deserve a special name. We call them
the {\it z--measures\/}.

When both $z,z'$ go to infinity, the expression \tht{1.1} has a limit
$$
M_{\infty,\infty,\th}^{(n)}(\la)=\frac{n!\,\th^n}{H(\la,\th)H'(\la,\th)}\,,
$$
which we call the {\it Plancherel measure\/} on $\Y_n$. The Plancherel measure
with $\th=1$ was considered in many works, see \cite{LS}, \cite{VK1},
\cite{VK3}, \cite{BDJ1}, \cite{BDJ2}, \cite{BDR}, \cite{BOO}, \cite{J1},
\cite{J2}, \cite{Ok3}.

The z--measures with $\th=1$ first originated in \cite{KOV} in connection with
the problem of harmonic analysis on the infinite symmetric group. The limits of
the measures $M_{z,z',1}^{(n)}$  as $n\to\infty$ govern the spectral
decomposition of the so--called generalized regular representations. The
z--measures with $\th=1$ and their limits were studied in detail in
\cite{P.I--P.V}, \cite{BO1--2}, \cite{Bor}, \cite{Ok2}.

Various special cases and degenerations of the z--measures with $\th=1$ also
arise in a number of problems not related to representation theory: see
\cite{J1}, \cite{J2}, \cite{TW}, \cite{GTW}, and our survey \cite{BO3}. Special
cases of z-measures with $\theta=1/2,2$ were considered in [AvM], [BR1], [BR2].

The z--measures with general $\th>0$ were first defined in \cite{Ke2} (see also
\cite{BO4} for another derivation). Besides $\th=1$, there exists one more
special value of the parameter $\th$ when the z--measures admit a
representation--theoretic interpretation: specifically, the case $\th=1/2$ is
related to a certain Gelfand pair associated with the infinite symmetric group.
No such interpretation exists for general $\th$. Nevertheless, introducing the
general parameter $\th$ seems to be a reasonable generalization. It is quite
similar to Heckman--Opdam's generalization of noncommutative spherical Fourier
analysis. Another motivation comes from comparison with log--gas (or random
matrix) models with general parameter $\be=2\th$.

The z--measures with different $n$ are related to each other by a {\it
coherency relation\/}, see Proposition 1.5 below. To state it, we need more
notation.

Let $P_\mu$ be the Jack symmetric
function with parameter $\th$ and index $\mu$ (see \cite{Ma2, VI.10}; note
that Macdonald uses $\alpha=\th^{-1}$ as the parameter). The simplest
case of Pieri's formula for the Jack functions reads as follows:
$$
P_\mu P_{(1)}=\sum_{\la:\, \la\searrow\mu}\dimth(\mu,\la)P_\la\,,
$$
where $\la\searrow\mu$ means that $\mu$ can be obtained from $\la$ by removing
one box, $\dimth(\mu,\la)$ are certain positive numbers. For the sake of
completeness, we give an explicit formula for $\dimth(\mu, \la)$, although we
will not use it in the sequel. We have
$$
\dimth(\mu,\la) = \prod_b \frac
{\big(a(b) + (l(b)+2)\th\big)\big(a(b) + 1 + l(b)\th\big)}
{\big(a(b)+(l(b)+1)\th\big)\big(a(b)+1+(l(b)+1)\th\big)}\,,
$$
where $b=(i,j)$ ranges over all boxes in the $j$th column of the diagram
$\mu$, provided that the new box $\la\setminus\mu$ belongs to the
$j$th column of $\la$, see \cite{Ma2, VI.10, VI.6},
$$
a(b)=a(i,j)=\mu_i-j,\quad
l(b)=l(i,j)=\mu'_j-i.
$$

For any $\mu\in\Y_{n-1}$ and $\la\in\Y_n$ set
$$
q_\th(\mu,\la)=\cases
\dfrac{H(\la,\th)}{n\,H(\mu,\th)}\,{\dimth(\mu,\la)},&
\lambda\searrow\mu,\\
0,&\text{otherwise}.\endcases
$$
For any $\lambda\in\Y_n$ we have
$$
\sum_{\mu\in\Y_{n-1}}q_\th(\mu,\la)=1.
$$
This relation readily follows from the Pieri formula for the Jack functions
above and the relation
$$
P_{(1)}^n=\sum_{\la\in\Y_n}\frac{n!}{H(\la,\th)}\,P_\la.
$$
Later on we will also use the notation
$$
C_\la=\frac{n!}{H(\la,\th)}\,P_\la.
$$

\proclaim{Proposition 1.5} For any $n=1,2,\dots$ and any $\mu\in\Y_{n-1}$ we
have
$$
M_{z,z',\th}^{(n-1)}(\mu)\equiv \sum_{\la\in\Y_n}
q_\th(\mu,\la)\,M_{z,z',\th}^{(n)}(\la),
$$
where we agree that $\Y_0=\{\varnothing\}$ and
$M_{z,z',\th}^{(0)}(\varnothing)=1$.
\endproclaim
\demo{Proof} See \cite{Ke2}, \cite{BO4}. \qed
\enddemo

It is convenient to view $\{q_\th(\mu,\la)\}$ as probabilities of a transition
from $\Y_n$ to $\Y_{n-1}$. Under this transition, the $n$th measure
$M_{z,z',\th}^{(n)}$ transforms into the $(n-1)$st measure $M_{z,z',\th}^{(n-
1)}$. Thus, the $n$th measure is a refinement of the $(n-1)$st one.

We are interested in the asymptotic behavior of the measures
$M_{z,z',\th}^{(n)}$ as $n\to\infty$. Since these measures live on different
spaces, we need to explain in what sense we understand the limit.

Let $\R^\infty=\R\times\R\times\cdots$ be the product of countably many copies
of the real line. We equip $\R^\infty$ with the product topology.
Set $\R^{2\infty}=\R^\infty\times\R^\infty$. Let $\Om$ be a subset of
$\R^{2\infty}$ consisting of pairs of sequences
$$
\al_1\ge\al_2\ge\dots\ge 0,\qquad \be_1\ge\be_2\ge\dots\ge0,
$$
subject to the condition
$$
\sum_{i=1}^\infty(\al_i+\be_i)\le 1.
$$
This is a metrizable compact topological space. Note that the subset
of $\Om$ with $\sum_i(\alpha_i+\be_i)=1$ is dense in $\Om$.

For any $n=1,2,\dots$, we define an embedding
$\iota_n:\Y_n\hookrightarrow \Om$ as follows. For any $\la\in\Y_n$, let
$d=d(\la)$ be the number of diagonal boxes of $\lambda$. Set
$$
\gather
a_i(\la)=\cases{\la_i-i+1/2}\,,&i\le d,\\
                             0,&i>d,
                 \endcases.
\qquad
b_i(\la)=\cases{\la'_i-i+1/2}\,,&i\le d,\\
                             0,&i>d,
                 \endcases
\\
\text{These are the {\it modified Frobenius coordinates} of $\la$ first
introduced in \cite{VK2}. Set}\\
\al_i(\la)=a_i(\la)/n,\qquad \be_i(\la)=b_i(\la)/n.
\endgather
$$
Note that $\sum_i(\al_i(\la)+\be_i(\la))=1$.
We define
$$
\iota_n(\la)=(\al_1(\la),\al_2(\la),\dots;
\be_1(\la),\be_2(\la),\dots).
$$
(In \cite{KOO}, the definition of $\iota_n$ was slightly different. This does
not affect, however, the following important claim, which is a special case of
one of the main results of \cite{KOO}. This follows, for instance, from Remark
1.7 below.)

\proclaim{Theorem 1.6} There exists a weak limit $\M_{z,z',\th}$ of the
pushforwards of the measures $M_{z,z',\th}^{(n)}$ under $\iota_n$:
$$
\M_{z,z',\th}=\operatorname{w-
lim}_{n\to\infty}\iota_n^*\left(M_{z,z',\th}^{(n)}\right).
$$
\endproclaim
\demo{Proof} See \cite{KOO}. Note that the claim holds for {\it any} system of
measures on $\Y_n$'s which satisfy the coherency relation of Proposition
1.5.\qed
\enddemo

\example{Remark 1.7} Consider the probability spaces
$(\Y_n,M_{z,z',\th}^{(n)})$ and consider the functions $\al_i(\,\cdot\,)$ and
$\be_i(\,\cdot\,)$ as random variables $\al_i^{(n)}$, $\be_i^{(n)}$ defined on
these spaces. Similarly, we view the coordinate functions $\al_i$, $\be_i$ on
$\Om$ as random variables defined on the probability space
$(\Om,M_{z,z',\th})$. Then Theorem 1.6 is equivalent to saying that for any
positive integers $m$, $l$,
$$
\{\al_1^{(n)},\dots,\al_m^{(n)},\be_1^{(n)},\dots,\be_l^{(n)}\} \overset
d\to\longrightarrow
\{\al_1,\dots,\al_m,\be_1,\dots,\be_l\},
$$
where $\overset d\to\longrightarrow$ denotes the convergence in distribution.
\endexample

Our main goal is to study the limit measures $\M_{z,z',\th}$.

The finite level measures $M_{z,z',\th}^{(n)}$ can be reconstructed from the
limit measure by means of an analog of the Poisson integral representation of
the harmonic functions. Let us briefly state this result. A more detailed
exposition can be found in \cite{KOO}.

Let $\La$ be the algebra of symmetric functions over $\R$. Following
\cite{KOO}, we will view the elements of $\La$ as
continuous functions on $\Om$. Namely, the values of the power sums $p_k$ are
defined by
$$
p_k(\al_1,\al_2,\dots;\be_1,\be_2,\dots)=\cases 1,&k=1,\\
\sum_i\left(\al_i^k+(-\th)^{k-1}\beta_i^k\right),&k\ge 2.
\endcases
$$
Since $\{p_k\}$ are free generators of the commutative algebra $\La$,
this defines an algebra homomorphism $\La\to C(\Om)$. In particular, the Jack
symmetric functions $\{P_\la\}$ can also be viewed as elements of
$C(\Om)$.

\proclaim{Theorem 1.8} For any $n=1,2,\dots$ and any $\la\in\Y_n$, we have
$$
M_{z,z',\th}^{(n)}(\la)=\frac{n!}{H(\la,\th)}\int_{\om=(\al,\be)\in\Om}
P_\la(\om)\M_{z,z',\th}(d\om).
$$
\endproclaim
\demo{Proof} See \cite{KOO}. Again, the claim holds for any
system of measures satisfying the coherency relation.\qed
\enddemo

Theorem 1.8 can also be interpreted in a different way, namely, as providing
the values of integrals of $\{P_\la\}$ with respect to the measure
$\M_{z,z',\th}$ on $\Om$. This set of integrals defines the limit measure
uniquely, because the functions $\{P_\la(\om)\}$ span a dense linear subspace
of $C(\Om)$. We view these integrals as ``moments'' of $\M_{z,z',\th}$.

Both descriptions of the measure $\M_{z,z',\th}$, as the weak limit (Theorem
1.6) and through the moments (Theorem 1.8), are rather abstract. Our goal is to
find yet another description which would allow us to obtain probabilistic
information about random points $\om=(\al_1,\al_2,\dots;\be_1,\be_2,\dots)$
distributed according to $\M_{z,z',\th}$.

It turns out to be very hard to compute directly the joint distribution
functions of finitely many $\al_i$'s or/and $\be_i$'s regarded as random
variables. Instead of that, we will focus on computing the {\it correlation
functions\/} of the measures $\M_{z,z',\th}$. Informally, the $n$th correlation
function of $\{\al_i\}$ measures the probability to find one $\alpha_i$ near
each of the $n$ given locations $x_1,\dots,x_n>0$:
$$
\multline
\rho_n(x_1,\dots,x_n)\\=\lim_{\Delta x_1,\dots,\Delta x_n\to +0}
\frac{\operatorname{Prob}\{\{\al_i\}\cap(x_j,x_j+\Delta x_j)\ne\varnothing\text{
for all  } j=1,\dots,n\}}{\Delta x_1\cdots \Delta x_n}\,.
\endmultline
$$

The correlation functions $\rho_n(x)$ should be viewed as densities of the {\it
correlation measures} $\rho_n(dx)$ with respect to the Lebesgue
measure $dx$. The knowledge of the correlation functions allows to evaluate
averages of the {\it additive} functionals on $\{\al_i\}$. Namely, for any
continuous function $F:\R^n_{>0}\to\C$ with compact support, we have
$$
\int_{\om=(\al;\be)\in\Om}\sum_{\Sb i_1,\dots,i_n\\ \text{pairwise
distinct}\endSb}F(\al_{i_1},\dots,\al_{i_n})\M_{z,z',\th}(d\om)=
\int_{\R^n_{>0}}F(x_1,\dots,x_n)\rho_n(dx).
$$
This equality can be viewed as a rigorous definition of $\rho_n(dx)$.
A detailed discussion of the correlation measures/functions can be
found in \cite{Len}, \cite{DVJ}.

Note that the correlation measure $\rho_n(dx)$ is supported by the simplex
$$
\{(x_1,\dots,x_n)\in(\R_{\ge 0})^n:x_1+\dots+x_n\le 1\}.
$$

More generally, one can similarly define {\it joint} correlations of
$\{\alpha_i\}$ and $\{\beta_i\}$. In the case $\th=1$ these joint correlation
functions have been computed in \cite{P.II}.

This definition of $\rho_n(dx)$ makes sense for an arbitrary probability
measure $M$ on $\Om$. Indeed, observe that for any point $\om=(\al,\be)\in\Om$,
we have the estimate
$$
\alpha_{m+1}<\frac 1m,\qquad m=1,2,\dots,
\tag 1.2
$$
which follows from the fact that $\al_1\ge \al_2\ge\dots$ and $\sum_i\al_i<1$.
For any nonnegative $F\in C_0((\R_{>0})^n)$, choose $m$ so large that $\supp
F\subset (\R_{\ge 1/m})^n$. Then in the above formula for $\langle
F,\rho_n\rangle$ the summands involving indices $i_k>m$ vanish. Thus,
the integrand is bounded by
$$
\sup F\cdot m(m-1)\cdots (m-n+1).
$$
This fact ensures the very existence of the correlation measures, see
\cite{Len}. It also implies a useful bound
$$
\rho_n((\R_{\ge 1/m})^n)\le m(m-1)\cdots(m-n+1)\le m^n,\qquad m=1,2,\dots\,.
\tag 1.3
$$

In the case $\th=1$ it was shown in \cite{P.II} that the expressions for
the correlation functions are substantially simplified by a one-dimensional
integral transform, see also \cite{P.III, P.V}, \cite{BO1-3}, \cite{Bor}. This
integral transform corresponds to a simple
modification of the initial measure on $\Om$. The modified measure for general
$\th$ is defined as follows.

Let us denote by $\wOm$ the set of triples
$\omega=(\al,\be,\delta)\in\R^{2\infty}\times\R_{\ge 0}$, where
$\al=(\al_1\ge\al_2\ge\dots\ge 0)$,  $\be=(\be_1\ge\be_2\ge\dots\ge 0)$,
$\delta\in \R_{\ge 0}$, and $ \sum_{i=1}^\infty(\al_i+\be_i)\le \delta. $ We
will also use the notation $\ga=\delta-\sum_i(\al_i+\be_i)\ge 0$.

Note that $\wOm$ is a locally compact space with respect to the topology induced
from the product topology on $\R^\infty\times\R_{\ge 0}$.
It is metrizable, the metric can be defined in the standard fashion:
$$
\dist(\om,\om')=|\delta-\delta'|+
\sum_i\frac{\min(|\al_i-\al_i'|,1)}{2^i}+
\sum_i\frac{\min(|\be_i-\be_i'|,1)}{2^i}\,.
$$
The subsets of $\wOm$ of the form $\{\om\in\wOm:\delta(\om)\le \const\}$ are
compact (here $\delta(\om)$ is the $\delta$--coordinate of $\om$).
The set $\{\om\in\wOm:\ga(\om)=0\}$ is everywhere dense in $\wOm$.

The space $\wOm$ is  homeomorphic to $\Omega\times \R_{\ge 0}$ modulo
contracting $\Omega\times\{0\}$ to a single point, the corresponding map looks
as follows
$$
\left(\left( \al,\be\right),\delta\right)\in \Omega\times\R_{\ge 0}\mapsto
(\delta\al,\delta\be,\delta)\in \wt\Omega .
$$
The modified measure $\wt\M_{z,z',\th}$ is the pushforward under this map of the
measure
$$
\M_{z,z',\th}\otimes \left(\frac{s^{t-1}}{\Ga(t)}\,e^{-s}ds\right)
$$
on $\Om\times\R_{\ge 0}$ (recall that $t=zz'/\th$).

The correlation measures/functions $\wt \rho_n$ of $\wt\M_{z,z',\th}$ are
defined in the same way as those of $\M_{z,z',\th}$. The definition of
$\wt\M_{z,z',\th}$ immediately implies that for any test function
$F\in\C_0((\R_{>0})^n)$,
$$
\langle F,\wt\rho_n\rangle=\int_0^\infty\frac{s^{t-1}e^{-s}}{\Ga(t)}\,\langle
F_s,\rho_n\rangle ds,
$$
where $F_s(x_1,\dots,x_n)=F(sx_1,\dots,sx_n)$.
In terms of the correlation functions (which may always be viewed as generalized
functions), we have
$$
\wt\rho_n(x_1,\ldots,x_n)=\int_0^\infty\frac{s^{t-1}e^{-s}}{\Gamma(t)}\,
\rho_n(x_1s^{-1},\ldots,x_ns^{-1})\frac {ds}{s^n}
\tag 1.4
$$
for any $n=1,2,\dots$\,.  The convergence of the integral follows from
\tht{1.3}.
This transform is easily reduced to the one--dimensional Laplace
transform along the rays $\{(\delta x_1,\dots,\delta x_n),\,\delta>0\}$. Hence,
it is invertible. The passage from $\M_{z,z',\th}$ to $\wt\M_{z,z',\th}$ is
called {\it lifting}.

The following proposition will be used in \S5.

\proclaim{Proposition 1.9} Let $F\in C_0((\R_{>0})^n)$ and $\delta\in\R_{>0}$.
Then the expression $\langle F_\delta,\wt\rho_n\rangle$, where
$F_\delta(x)=F(\delta\cdot x)$ as above, is a real--analytic function of
$\delta$.
\endproclaim

\demo{Proof} We have
$$
\langle F_\delta,\wt\rho_n\rangle=\int_0^\infty\frac{s^{t-1}e^{-
s}}{\Ga(t)}\,\langle
F_{s\delta},\rho_n\rangle ds=\delta^{-t}\int_0^\infty
\frac{s^{t-1}e^{-s/\delta}}{\Ga(t)}\,\langle F_s,\rho_n\rangle ds.
$$
Pick $\epsilon>0$ such that $\supp F\subset (\R_{\ge \epsilon})^n$.
The claim follows from the following two facts:

1. $\langle F_s,\rho_n\rangle$ vanishes for $s<\epsilon$.

2. $\langle F_s,\rho_n\rangle$ has at most polynomial growth in $s$
when $s\to\infty$.

The vanishing follows from the fact that $\supp
\rho_n\subset\{\sum_{i=1}^nx_i\le 1\}$.

For the second fact, observe that by \tht{1.3} we have
$$
|\langle F_s,\rho_n\rangle|\le \sup |F|\cdot \rho_n((\R_{\ge s^{-
1}\epsilon})^n)\le \sup |F|\cdot ([s\epsilon^{-1}]+1)^n.\qed
$$
\enddemo

\example{Remark 1.10} In the case when $(z,z')$ belong to the degenerate series
(see the definition above), the measures $M_{z,z',\th}^{(n)}$ and their limit
$\M_{z,z',\th}$ were studied by Kerov \cite{Ke1}. To be concrete, assume that
$z=m\th$, $m=1,2,\dots$, and $z'>(m-1)\th$. Then the limit measure
$\M_{z,z',\th}$ is concentrated on the $(m-1)$-dimensional face
$$
\left\{(\al,\be)\in\Om:
\al_1+\dots+\al_m=1,\,\al_{m+1}=\al_{m+2}=\dots=\be_1=\be_2=\dots=0 \right\}.
$$
Its density with respect to the Lebesgue measure on this simplex is equal to
$$
\const\cdot(\al_1\cdots\al_m)^{z'-(m-1)\th-1}\cdot\prod_{1\le i<j\le m}|\al_i-
\al_j|^{2\th}.
\tag 1.5
$$
The lifting $\M_{z,z',\th}$ of this measure lives on $(\R_{\ge 0})^m$ and has
the density (with respect to the Lebesgue measure) equal to
$$
\const\cdot(\al_1\cdots\al_m)^{z'-(m-1)\th-1}\cdot e^{-\al_1-\dots-
\al_m}\cdot\prod_{1\le i<j\le m}|\al_i-\al_j|^{2\th}.
\tag 1.6
$$
This is the distribution function for the $m$-particle Laguerre ensemble, see
\cite{F1}, \cite{F2}.
\endexample

\head 2. Averages of $E_\th(\,\cdot\,;u_1)\cdots E_\th(\,\cdot\,;u_l)$ as
hypergeometric functions
\endhead
Set
$$
E_\th(\om;u)=
e^{\ga/u}\frac{\prod_{i=1}^\infty(1+\al_i/u)}{\prod_{i=1}^\infty(1-
\th\be_i/u)^{1/\th}}\,,\qquad
\om\in\wOm,\ \ u\in \C\setminus\R_{\ge 0}.
$$

Let us comment on this definition. Consider the algebra homomorphism
$\Lambda\to C(\widetilde\Omega)$ defined on the power sums by
$$
p_1(\omega)=\delta; \qquad p_k(\omega)=\sum
\alpha_i^k+(-\theta)^{k-1}\sum\beta_i^k, \quad k\ge2.
$$
This is an algebra embedding generalizing the homomorphism $\Lambda\to
C(\Omega)$ as defined in section 1. Then $E_\theta(\omega;u)$ is nothing but
the image  of the generating function $\sum e_k u^{-k}$, where $e_k\in\Lambda$
are the elementary symmetric functions.

We view $E_\th(\om;u)$ as the analog of the characteristic polynomial of a
matrix, the roles of ``eigenvalues'' are played by $\alpha_i$'s and $\beta_i$'s.

One can show that for any $u\in\C\setminus \R_{\ge 0}$, the function
$E_\th(\,\cdot\,;u)$ is a continuous function on $\wOm$, cf. \cite{KOO}, and for
any $\om\in\wOm$, $E_\th(\om;\,\cdot\,)$ is a
holomorphic function on $\C\setminus\R_{\ge 0}$.

Observe that $E_\th$ is homogeneous of degree 0:
$$
E_\th(s\cdot\om;s\cdot u)=E_\th(\om;u),\qquad s>0.
$$

We will also consider $E_\th(\,\cdot\,;u)$ as a function on $\Om$.\footnote{In
what follows we view $\Om$ as a subset of $\wOm$ defined by the condition
$\delta=1$.} Then the domain of $u$ can be expanded to $\C\setminus[0,\th]$.

The goal of this section is to express the averages ($l=1,2,\dots$)
$$
\int_\Om E_\th(\,\cdot\,;u_1)\cdots
E_\th(\,\cdot\,;u_l)\M_{z,z',\th}(d\om),\qquad
\int_{\wOm} E_\th(\,\cdot\,;u_1)\cdots E_\th(\,\cdot\,;u_l)
\wt\M_{z,z',\th}(d\om)
$$
in terms of multivariate hypergeometric functions.

 Recall that in the previous section we introduced the renormalized Jack
polynomials $C_\la=C_\la^{(\nu)}(x)$. Here we deliberately included the
parameter $\nu$ in the notation of the Jack polynomials. In \S1 this parameter
was equal to $\th$, and in this section we will need $\nu=\tth$.

For $a,b,c\in\C$, $c\ne 0,-1,-2,\dots$, set
$$
{{}_2\wh F_1^{(\nu)}}(a,b;c;x)=\sum_{\Sb \la\in\Y\\ \ell(\la)\le l\endSb}
\frac{(a)_{\la,\nu}(b)_{\la,\nu}}{(c)_{|\la|}|\la|!}\,C^{(\nu)}_\la(x),\quad
x=(x_1,\dots,x_l).
$$
Note that the normalized series
$$
\frac{{}_2\wh F_1^{(\nu)}(a,b;c;x)}{\Ga(c)}=\sum_{\Sb \la\in\Y\\
\ell(\la)\le l\endSb}
\frac{(a)_{\la,\nu}(b)_{\la,\nu}}{\Ga(c+|\la|)\,|\la|!}\,C^{(\nu)}_\la(x),\quad
x=(x_1,\dots,x_l)
$$
makes sense for any $c\in\C$.

When $l=1$, the definition of ${{}_2\wh F_1^{(\nu)}}(a,b;c;x)$ above coincides
with that of the classical Gauss hypergeometric function. When $l>1$ our series
differs from the standard multivariate generalization of the Gauss function,
see \cite{Mu}, \cite{Ma1}, \cite{Ko},
\cite{FK}, \cite{Y}. Indeed, in the standard definition one has
$(c)_{\la,\nu}$ instead of $(c)_{|\la|}$ in the denominator. However, our
function ${{}_2\wh F_1^{(\nu)}}(a,b;c;x)$ shares many properties of the
standard hypergeometric functions.

\proclaim{Proposition 2.1} {\rm (i)} The defining series for ${{}_2\wh
F_1^{(\nu)}}(a,b;c;x)$ converges in the polydisk $\{|x_1|<1,\dots,|x_l|<1\}$
and defines a holomorphic function in this domain.

{\rm (ii)} ${}_2\wh F_1^{(\nu)}(a,b;c;x)/\Ga(c)$ is an entire function in the
parameters $(a,b,c)\in\C^3$. As a function in $x$, it can be analytically
continued to a domain in $\C^l$ containing the tube
$\{(x_1,\dots,x_l)\in\C^l:\, \Re x_i<0,\, i=1,\dots,l\}$.

{\rm (iii)} As $x_1,\dots,x_l\to -\infty$ inside $\R$, $|{}_2\wh
F_1^{(\nu)}(a,b;c;x)|$ has at most polynomial growth in $x$.
\endproclaim
\demo{Idea of proof} (i) Compare the series ${}_2\wh F_1^{(\nu)}(a,b;c;x)$ with
the series
$$
{}_1F_0^{(\nu)}(a;x)=\sum_{\Sb \la\in\Y\\ \ell(\la)\le l\endSb}
\frac{(a)_{\la,\nu}}{|\la|!}\,C^{(\nu)}_\la(x),\quad x=(x_1,\dots,x_l).
$$
By virtue of the well--known binomial theorem (see, e.g.,
\cite{Ma1}, \cite{OO})
$$
{}_1F_0^{(\nu)}(a;x)=\prod_{i=1}^l(1-x_i)^{-a},
$$
which implies that the latter series converges in the polydisk in question.
Since the ratio $(b)_{\la,\nu}/(c)_{|\la|}$ has at most polynomial growth in
$|\la|$, the former series also converges in the same polydisk.

(ii) An argument is given below after Proposition 2.2.

(iii) This can be derived from a Mellin--Barnes integral representation for
${}_2\wh F_1^{(\nu)}(a,b;c;x)$, which will be given elsewhere. \qed
\enddemo

Consider the multivariate hypergeometric function of type (1,0) in two sets of
variables $x=(x_1,\dots,x_l)$ and $y=(y_1,\dots,y_l)$:
$$
{}_1\F_0^{(\nu)}(a;x,y)=\sum_{\Sb \la\in\Y\\ \ell(\la)\le
l\endSb}\frac{(a)_{\la,\nu}}{|\la|!} \,\frac{C_\la^{(\nu)}(x)
C_\la^{(\nu)}(y)}{C_\la^{(\nu)}(1^l)}\,,\qquad a\in\C,\ \nu>0,
$$
see \cite{Ma1}, \cite{Y, (37)}. When $\nu=1/2,\,1,\,2$, this
function admits a simple matrix integral representation. For instance, in the
case $\nu=1/2$
$$
{}_1\F_0^{(\nu)}(a;x,y)=\int_{U\in O(l))}\det(1-XUYU^{-1})^{-a}dU,
$$
where $O(l)$ is the group of $l\times l$ orthogonal matrices, $dU$ is the
normalized Haar measure on $O(l)$, and  $X$ and $Y$ stand for the diagonal
matrices with diagonal entries $(x_i)$ and $(y_i)$.

The next statement gives an Euler-type integral representation of ${}_2\wh
F_1^{(\nu)}(a,b;c;x)$ in terms of ${}_1\F_0^{(\nu)}$. For the three particular
values of the parameter, $\nu=1/2,\, 1,\, 2$, it can be written as a matrix
integral involving elementary functions only.

\proclaim{Proposition 2.2} For any $\nu>0$, assume that $\Re b>(l-1)\nu$, $\Re
c>l\,\Re b$. Then
$$
\gathered
\frac{{}_2\wh F_1^{(\nu)}(a,b;c;x)}{\Ga(c)}=
\frac1{\Gamma(c-lb)}\prod_{j=1}^l\frac{\Ga(\nu+1)}{\Ga(b-(j- 1)\nu)\Ga(j\nu+1)}
\\ \times
\int\limits_{\Sb \tau_1,\dots,\tau_l>0\\ \sum_i\tau_i< 1\endSb} \prod_{i=1}^l
\tau_i^{b-(l-1)\nu-1}\,\left(1-\sum_{i=1}^l\tau_i\right)^{c-lb-1} \prod_{1\le
i<j\le l}|\tau_i-\tau_j|^{2\nu}{}_1\F_0^{(\nu)} (a;x,\tau)d\tau.
\endgathered \tag2.1
$$
\endproclaim

\demo{Proof} We use the following integral representation of the ratio
$(b)_{\la,\nu}/\Ga(c+|\la|)$
$$
\gathered \frac{(b)_{\la,\nu}}{\Ga(c+|\la|)}=
\frac1{\Gamma(c-lb)}\prod_{j=1}^l\frac{\Ga(\nu+1)}{\Ga(b-(j- 1)\nu)\Ga(j\nu+1)}
\\ \times\int\limits_{\Sb \tau_1,\dots,\tau_l>0\\
\sum_i\tau_i< 1\endSb} \prod_{i=1}^l
\tau_i^{b-(l-1)\nu-1}\,\left(1-\sum_{i=1}^l\tau_i\right)^{c-lb-1} \prod_{1\le
i<j\le
l}|\tau_i-\tau_j|^{2\nu}\,\frac{C^{(\nu)}_\la(\tau)}{C^{(\nu)}_\la(1^l)}\,d\tau.
\endgathered \tag2.2
$$
A derivation of \tht{2.2} is given in \cite{Ma2, ch. VI, \S10, Example 7 (b)}.
Multiplying both sides of \tht{2.2} by
$$
\frac{(a)_{\la,\nu}}{|\la|!}\,C^{(\nu)}_\la(x),
$$
taking the sum over $\la$ and interchanging summation and integration, one
obtains the required equality. \qed
\enddemo

Note that the $l$--dimensional integral \tht{2.2} is a consequence of the
following integral over an $(l-1)$--dimensional simplex
$$
\gathered \int\limits_{\Sb t_1+\dots+t_l=1\\ t_1,\dots,t_l\ge0\endSb}
\prod_{j=1}^l t_j^{A-1}\, \prod_{1\le i<j\le l}|t_i-t_j|^{2\nu}\,
\frac{C^{(\nu)}_\la(t_1,\dots,t_l)}{C^{(\nu)}_\la(1^l)}\,dt\\
=\frac1{\Ga(|\la|+Al+l(l-1)\nu)}\, \prod_{j=1}^l
\frac{\Ga(\la_j+A+(l-j)\nu)\Ga(j\nu+1)}{\Ga(\nu+1)}\,,
\endgathered \tag2.3
$$
where $\Re A>0$ and $dt$ is Lebesgue measure on the simplex.

The integral \tht{2.2} can be derived from the integral \tht{2.3} as follows:
Set $\tau=ts$, where $s=\sum\tau_j$. Since the integrand of \tht{2.2} is a
homogeneous function, the integral splits into the product of an
$(l-1)$--dimensional integral over $t$ (which is the integral \tht{2.3} with
$A=b-(l-1)\nu$) and a one--dimensional beta--integral over $s$.

As for the integral \tht{2.3}, it is a simplex version of the generalized
Selberg integral over the unit cube $[0,1]^l$, see \cite{Ma2, ch. VI, \S10,
example 7}. Once one knows the integral over the cube, it is easy to pass to
the simplex. On the other hand, the integral \tht{2.3} can be obtained directly
by making use of degenerate z--measures, see Kerov \cite{Ke1, \S12}.

\demo{Sketch of proof of Proposition 2.1 (ii)} Our argument is based on the
Euler--type integral representation \tht{2.1}. We will prove that the integral
\tht{2.1}, as a function in $x$, can be continued to the tube $\{x\in\C^l:\,
\Re x_i<1/2, \, i=1,\dots,l\}$. This result is not optimal: when
$\nu=1/2,\,1,\,2$, use of the matrix integral representation for ${}_1\Cal
F_0^{(\nu)}(x,\tau)$ allows one to extend the domain to the tube $\{x\in\C^l:\,
\Re x_i<1, \, i=1,\dots,l\}$ (cf. \cite{FK, Prop. XV.3.3}).

Assume first $\Re b>(l-1)\nu$ and $\Re(c-lb)>0$ so that the integrand in
\tht{2.1} is an integrable function (then we will explain how to get rid of
these restrictions).

The idea is to apply the transformation formula
$$
{}_1\Cal F_0^{(\nu)}(x,y)=\prod_{j=1}^l (1-x_j)^{-a}\cdot{}_1\Cal
F_0^{(\nu)}\left(\frac{x}{x-1}\,,\,1-y\right),\tag2.4
$$
established in Macdonald \cite{Ma1, section 6}. Here we abbreviate
$$
\frac{x}{x-1}=\left(\frac{x_1}{x_1-1}\,, \,\dots,\,\frac{x_l}{x_l-1}\right),
\qquad 1-y=(1-y_1,\dots,1-y_l).
$$
When $\nu=1/2,\,1,\,2$, the transformation \tht{2.4} is immediate from the
matrix integral representation of ${}_1\Cal F_0^{(\nu)}$. But in the general
case, when we dispose of the series expansion only, \tht{2.4} is not evident.
(Note that Macdonald's argument uses some properties of generalized binomial
coefficients and Jack polynomials, admitted as conjectures. But nowadays these
are well-established facts.)

Since $\zeta\mapsto \zeta(\zeta-1)^{-1}$ transforms the half--plane
$\Re\zeta<1/2$ into the unit disk $|\zeta|<1$, the transformation \tht{2.4} can
be used to correctly define ${}_1\Cal F_0^{(\nu)}(x,y)$ when $x$ ranges over
the tube $\Re x_i<1/2$ and $y=\tau$.

Thus, we checked that the required analytic continuation in $x$ exists under an
additional restriction on the parameters $b,c$. Let us show how to get rid of
this restriction. Take a large constant $C>0$ and assume first that $\Re c>lC$.
Then, as a function in $(a,b)$, our integral admits a continuation to the tube
domain $\{(a,b)\in\C^2:\, \Re a<C,\, (l-1)\nu<\Re b<C\}$. By virtue of symmetry
$a\leftrightarrow b$, the same holds for the tube $\{(a,b)\in\C^2:\,
(l-1)\nu<\Re a<C,\, \Re b<C\}$. Applying a general theorem about ``forced''
analytic continuation on tube domains (see, e.g., \cite{H, Theorem
2.5.10}) we obtain a continuation to the tube $\{(a,b)\in\C^2:\, \Re a<C,\, \Re
b<C\}$. Finally, to remove the restriction on $c$, we use the relation
$$
(c-1+\Cal D)\left(\frac{{}_2\wh F_1^{(\nu)}(a,b;c;x)}{\Ga(c)}\right)=
\frac{{}_2\wh F_1^{(\nu)}(a,b;c-1;x)}{\Ga(c-1)}\,,
$$
where $\Cal D$ is the Euler operator,
$$
\Cal D=\sum_{j=1}^l x_j\,\frac{\partial}{\partial x_j}\,,
$$
which follows from the initial series expansion for ${}_2\wh
F_1^{(\nu)}(a,b;c;x)/\Ga(c)$ and the fact that $C^{(\nu)}_\la(x)$ is a
homogeneous function of degree $|\la|$. \qed
\enddemo

We return to our main subject.

\proclaim{Theorem 2.3} Let $l=1,2,\dots$, and let $\Re u_i<0$, \,
$i=1,\dots,l$. Then
$$
\gather
\int_\Om E_\th(\om;u_1)\cdots E_\th(\om;u_l)\M_{z,z',\th}(d\om)={{}_2\wh
F_1^{(\nu)}}(a,b;c;\th/u),
\endgather
$$
where
$$
\gather
\nu=\tth,\quad \th/u=(\th/u_1,\dots,\th/u_l),\\ a=-z\tth,\quad b=-z'\tth,\quad
c=zz'\tth.
\endgather
$$
\endproclaim
\demo{Proof} Observe that $\Om$ is compact and $E_\th(\,\cdot\,;u)\in C(\Om)$,
thus, the integral is well-defined. Since both sides of the equality in
question are holomorphic in $u_1,\dots,u_l$, we may assume that $|u_i|\gg 0$.

The dual Cauchy identity for the ordinary Jack polynomials (see
\cite{Ma2, Ch. VI, (5.4)}) implies the expansion
$$
E_\th(\om;u_1)\dots E_\th(\om;u_l)=\sum_{\la:\,\ell(\la)\le l}
P_{\la'}^{(\th)}(\om)P_{\la}^{(\tth)}(u_1^{-1},\dots,u_l^{-1}), \quad \om\in\Om.
$$
 Let us integrate the series over $\Om$ termwise. By Theorem 1.8 and \tht{1.1},
for any $\la\in\Y_n$,
$$
\int\limits_{\om\in\Om}
P_{\la'}^{(\th)}(\om)\M_{z,z',\th}(d\om)=\frac{H(\la',\th)}{n!}\,M_{z,z',\th}^{(
n)}(\la')=\frac{(z)_{\la',\th}(z')_{\la',\th}}{(t)_nH'(\la',\th)}\,.
$$
An easy computation shows that
$$
\gather
(z)_{\la',\th}(z')_{\la',\th}=\th^{2n}(-z\tth)_{\la,\tth}
(-z'\tth)_{\la,\tth},\\
H'(\la',\th)=\th^n H(\la,\tth).
\endgather
$$
Since $C_\la^{(\tth)}=n!P_\la^{(\tth)}/H(\la,\tth)$, the claim follows.\qed
\enddemo

We would like to obtain an analog of Theorem 2.3 when $\Om$ is replaced by
$\wOm$ and $\M_{z,z',\th}$ is replaced by the lifted measure $\wt
\M_{z,z',\th}$. By definition of $\wt\M_{z,z',\th}$ and Fubini's theorem, we
have
$$
\multline
\int_{\wOm} E_\th(\om;u_1)\dots E_\th(\om;u_l) \wt\M_{z,z',\th}(d\om)\\=
\int_0^\infty\frac{s^{t-1}}{\Ga(t)}\,e^{-s}\left(\int_\Om
E_\th(s\cdot\om;u_1)\dots E_\th(s\cdot\om;u_l)\M_{z,z',\th}(d\om)\right)ds,
\endmultline
$$
provided that the integral exists. By the 0-homogeneity property of
$E_\th(\om;u)$ we can rewrite the integral as
$$
\int_0^\infty\frac{s^{t-1}}{\Ga(t)}\,e^{-s}\left(\int_\Om E_\th(\om;u_1/s)\dots
E_\th(\om; u_l/s)\M_{z,z',\th}(d\om)\right)ds.
$$
Hence, by Theorem 2.3, this equals
$$
\int_0^\infty\frac{s^{t-1}}{\Ga(t)}\,e^{-s}
{{}_2\wh F_1^{(\nu)}}(a,b;c;\,s\th/u)ds.
$$
Recall that $t=c=zz'\th^{-1}$.
This computation suggests the following definition.

For $a,b\in\C$, set
$$
{}_2F_0^{(\nu)}(a,b;x)=\int_0^{\infty}\frac{s^{c-1}}{\Ga(c)}\,e^{-s} {{}_2\wh
F_1^{(\nu)}}(a,b;c;\,s\cdot x)ds,\qquad c>0,\quad x=(x_1,\dots,x_l). \tag 2.5
$$
As will be shown below, see Proposition 2.4, the right--hand side does not
depends on the choice of $c$. By Proposition 2.1(iii), the integral above makes
sense at least when $x_1,\dots,x_l<0$.

The notation ${}_2F_0^{(\nu)}$ is justified by the following formal argument:
applying the integral transform to the series expansion of ${}_2\wh
F_1^{(\nu)}$ we obtain the series
$$
{}_2F_0^{(\nu)}(a,b;x)=\sum_{\Sb \la\in\Y\\ \ell(\la)\le l\endSb}
\frac{(a)_{\la,\nu}(b)_{\la,\nu}}{|\la|!}\,C^{(\nu)}_\la(x).
$$
Note that the series in the right--hand side does not depend on $c$. However,
if $a,b$ are not equal to $0,-1,-2,\dots$, this series is {\it everywhere
divergent\/} (except the origin).\footnote{If one of the parameters $a$ and $b$
is equal to $0,-1,-2,\dots$, then the series terminates and defines a
polynomial, which can also be written through ${}_1F_1$ series, see Remark 2.6
below.} Such phenomenon is well known already in the classical one--dimensional
case, see \cite{Er, section 5.1}. Our definition is one possibility to
circumvent this difficulty in making sense of ${}_2F_0$.

\proclaim{Proposition 2.4}
 For any $\nu>0$, assume that $\Re b>(l-1)\nu$. Then
$$
\gather
{}_2 F_0^{(\nu)}(a,b;x)=
\prod_{j=1}^l\frac{\Ga(\nu+1)}{\Ga(b-(j-1)\nu)\Ga(j\nu+1)}
\\ \times
\int\limits_{\tau_1,\dots,\tau_l>0}
\prod_{i=1}^l \tau_i^{b-\nu(l-1)-1}e^{-\tau_i}\prod_{1\le i<j\le l}|\tau_i-
\tau_j|^{2\nu}{}_1\F_0^{(\nu)}
(a;x;\tau)d\tau.
\endgather
$$
\endproclaim
\demo{Proof} By the homogeneity, ${}_1\F_0^{(\nu)}(a;\,s\cdot x;\tau)=
{}_1\F_0^{(\nu)}(a;x;\,s\cdot\tau)$. Using Theorem 2.3 and changing the
variables $s\cdot\tau_i=\sigma_i$, we obtain
$$
\gathered
{}_2 F_0^{(\nu)}(a,b;x)=\prod_{j=1}^l\frac{\Ga(\nu+1)}{\Ga(b-(j-
1)\nu)\Ga(j\nu+1)}\int\limits_{\sigma_1,\dots,\sigma_l>0}\prod_{i=1}^l
\sigma_i^{b-\nu(l-1)-1}\\ \times\prod_{1\le i<j\le l}|\sigma_i-
\sigma_j|^{2\nu}{}_1\F_0^{(\nu)}
(a;x;\sigma)\left(\int_{0}^\infty\frac{\left(s-\sum_i\sigma_i\right)^{d-
1}}{\Ga(d)}\,ds\right)d\sigma,
\endgathered
$$
which immediately gives the desired formula.\qed
\enddemo

Similarly to the one--dimensional case, the function ${}_2 F_0^{(\nu)}(a,b;x)$
can be analytically continued to tube $\{x\in\C^l:\, \Re x_i<0,\,
i=1,\dots,l\}$. The divergent series for ${}_2F_0$ given above is, in fact, the
asymptotic expansion of ${}_2F_0$ near $x=0$.

When $l=1$, we have ${}_1\F_0^{(\nu)}(a;x;\tau)=(1-x\tau)^{-a}$, so that the
dependence on $\nu$ disappears and Proposition 2.4 takes the form
$$
{}_2F_0(a,b;x)=\frac 1{\Ga(b)}\int_0^\infty \tau^{b-1}(1-x\tau)^{-a}e^{-
\tau}d\tau.
$$
This is equivalent to the classical integral representation for the Whittaker
function $\Psi$, see \cite{Er, 6.5(2)} (note that ${}_2F_0$ and
Whittaker's $\Psi$ are essentially the same functions, see \cite{Er,
6.6(3)}).

Again, when $\nu=1/2,\,1,\,2$ (and $l$ is arbitrary), we dispose of a matrix
integral representation for ${}_2F_0(a,b;x)$. In the case $\nu=1/2$, the
integral was studied in detail in \cite{MP1}, \cite{MP2}.

\proclaim{Theorem 2.5} For any $l=1,2,\dots$, and $u_1,\dots,u_l<0$, the
product $E_\th(\om;u_1)\cdots E_\th(\om;u_l)$ as a function on $\wOm$ is
integrable with respect to the measure $\wt\M_{z,z',\th}$ on $\wOm$, and
$$
\int_{\wOm} E_\th(\om;u_1)\cdots
E_\th(\om;u_l)\wt\M_{z,z',\th}(d\om)={{}_2F_0^{(\nu)}}(a,b;\th/u), \tag2.6
$$
where
$$
\gather
\nu=\tth,\quad \th/u=(\th/u_1,\dots,\th/u_l),\\ a=-z\tth,\quad b=-z'\tth.
\endgather
$$
\endproclaim
\demo{Proof}
If we take the integrability for granted then the statement follows from Theorem
2.3 and definition of ${}_2F_0$ as was explained above. To prove the
integrability, it suffices to show that
$$
\multline
\int_{\wOm} \left|E_\th(\om;u_1)\cdots
E_\th(\om;u_l)\right|^2\wt\M_{z,z',\th}(d\om)\\=
\int_0^\infty\frac{s^{t-1}}{\Ga(t)}\,e^{-s}\left(\int_\Om
|E_\th(s\cdot\om;u_1)\dots E_\th(s\cdot\om;u_l)|^2\M_{z,z',\th}(d\om)\right)ds
<\infty,
\endmultline
$$
because the total measure of the whole space $\wOm$ is finite. By Theorem 2.3,
the integral over $\Om$ equals
$$
{}_2\wh F_0(a,b;c;s\cdot\th/u,s\cdot\th/u),
$$
which grows at most polynomially as $s\to\infty$. \qed
\enddemo

\example{Remark 2.6} Assume, as in Remark 1.10, that $z=m\th$, $m=1,2,\dots$,
so that $a=-m$ in Theorem 2.5 above. In this case $E_\th(\om;u)$ reduces to
$$
E_\th(\om;u)=u^{-m}\prod_{i=1}^m(u+\al_i).
$$
Then the integral in the left--hand side of \tht{2.6} takes the form
$$
\multline
\const\cdot(u_1\cdots u_l)^{-m}\\
\times
\int_{(\R_{\ge 0})^m}\prod_{j=1}^l\prod_{i=1}^m(u_j+\al_i)\cdot\prod_{1\le
i<j\le m}|\al_i-\al_j|^{2\th}\cdot\prod_{i=1}^m\al_i^{z'-(m-1)\th-1}e^{-\al_i}
d\al_i.
\endmultline
$$

On the other hand, one can prove the general identity: for $m=1,2,\dots$,
$$
\gather
{}_2F_0^{(\nu)}(-m,b;x_1^{-1},\cdots,x_l^{-1})
=\prod_{i=1}^l (b-(i-1)\nu)_m \cdot (x_1\cdots x_l)^{-m}\\ \times
{}_1F_1^{(\nu)}(-m;-b-m+1+(l-1)\nu;-x_1,\dots,-x_l).
\endgather
$$
(Note that the series for ${}_1F_1$ in the right--hand side terminates.)

Thus, \tht{2.6} turns into (using the notation $A=z'-(m-1)\th>0$)
$$
\gather
\int_{(\R_{\ge 0})^m}\prod_{j=1}^l\prod_{i=1}^m(u_j+\al_i)\cdot\prod_{1\le
i<j\le m}|\al_i-\al_j|^{2\th}\cdot\prod_{i=1}^m\al_i^{A-1}e^{-\al_i} d\al_i\\
=\const \cdot
{}_1F_1^{(1/\th)}\left(-m;\frac{A+l-1}\th;-\frac{u_1}{\th},\dots,-
\frac{u_l}{\th}\right).
\endgather
$$
This agrees with the results of \cite{Ka} and \cite{BF}.
\endexample

\example{Remark 2.7} The formula
$$
{}_1\Cal F_0^{(\nu)}(a; \underbrace{x,\dots, x}_{\text{$l$ times}};
\tau_1,\dots,\tau_l)=\prod_{i=1}^l (1-x\tau_1)^{-a}
$$
shows that the integral representations of Propositions 2.2 and 2.4 in the case
when $x_1=\dots=x_l=x$ involve elementary functions only.
\endexample

\head 3. Lattice correlation functions
\endhead

The lifting transform introduced at the end of \S1 has a natural discrete
counterpart.
Starting with probability measures $M_{z,z',\th}^{(n)}$ on $\Y_n$,
$n=0,1,\dots$, we define a probability measure $\wt M_{z,z',\th;\xi}$
on the set $\Y=\Y_0\sqcup\Y_1\sqcup\Y_2\sqcup\dots$ of all Young diagrams with
an additional parameter $\xi\in(0,1)$ by
$$
\wt M_{z,z',\th;\xi}(\la)=(1-\xi)^t\,\frac{(t)_n}{n!}\xi^n\, \cdot
M^{(n)}_{z,z',\th}(\la),\qquad n=|\la|.
$$
That is, we mix the measures on $\Y_n$'s using the {\it negative binomial
distribution} $\{(1-\xi)^t\,\frac{(t)_n}{n!}\xi^n\}$ on nonnegative integers
$n$.

In the particular case $\th=1$, these mixed measures on $\Y$ were introduced in
\cite{BO2}. They are a special case of Okounkov's Schur measures defined in
\cite{Ok4}. For general $\th>0$, the measures $\wt M_{z,z',\th;\xi}$ are a
special case of ``Jack measures'' --- a natural extension of Okounkov's
concept.

In the next section we will show that the lifted measure $\wt \M_{z,z',\th}$ on
$\wOm$ can be obtained as a limit of the discrete mixed measures $\wt
M_{z,z',\th;\xi}$ as $\xi\nearrow 1$.

For the rest of this section we assume that $\th$ is a positive integer:
$\th=1,2,3,\dots$\,.

To a Young diagram $\la$ we assign a semiinfinite point
configuration $\LL=\LL(\la)$ on $\Z$, as follows
$$
\LL=\{l_1,l_2,\dots\}, \qquad l_i:=\la_i-i\th.
$$
In particular,
$$
\LL(\varnothing)=\{\lv_1,\lv_2,\lv_3,\dots\}=\{-\th,-2\th,-3\th,\dots\}.
$$

\proclaim{Proposition 3.1} A sequence of integers $\LL=(l_1,l_2,\dots)$ is of
the form $\LL=\LL(\la)$ for some Young diagram $\la$ if and only if the
following conditions hold:

{\rm(i)} $l_i-l_{i+1}\ge\th$ for all $i$.

{\rm(ii)} If $i$ is large enough then $l_i-l_{i+1}=\th$.

{\rm(iii)} The stable value of the quantity $l_i+i\th$, whose existence
follows from {(ii)}, equals 0.
\endproclaim

\demo{Proof} The above conditions are clearly necessary. Let us check
that they are sufficient. Set $\la_i=l_i+i\th$.  Condition (i)
implies that $\la_i\ge\la_{i+1}$. Conditions (ii) and (iii) imply
that $\la_i=0$ for all $i$ large enough. Hence
$\la=(\la_1,\la_2,\dots)$ is a partition. \qed
\enddemo

Let $\LL$ satisfy the conditions (i)--(iii) from Proposition 3.1.  Let
$a\in\LL$. If one removes $a$ from $\LL$
then the new configuration $\LL\setminus\{a\}$ will satisfy
(i) and (ii) but not (iii). Indeed, in $\LL\setminus\{a\}$,
the stable value of the quantity $l_i+i\th$ will be equal to $-\th$,
not 0. To compensate, we shift the whole $\LL\setminus\{a\}$  by $\th$ (that is,
we add $\th$ to all members of the sequence).
Then (i) and (ii) remain intact while the stable value in (iii)
becomes equal to 0, as required. Let us denote the resulting configuration by
$\Cal D_a(\LL)$.

Observe that $\Cal D_a(\LL)$ does not intersect
$\{a+1,\dots,a+2\th-1\}$. Conversely, any configuration that
satisfies this property together with (i)--(iii) has the form $\Cal
D_a(\LL)$ for a certain configuration $\LL$ satisfying (i)--(iii).

One could also define the inverse operation: given a
configuration satisfying (i)--(iii) and not intersecting
$\{a+1,\dots,a+2\th-1\}$, we add to it the point $a+\th$ and then
shift all the points  by $-\th$.

We use the same symbol $\Cal D_a$ to denote the corresponding operation
on Young diagrams. In diagram notation, this operation looks as follows. Given
$\la\in\Y$, let $j$ be such that
$\la_j-j\th=a$, which is equivalent to $l_j=a$ (if there is no such
$j$ then the operation is not defined). Then
$$
\Cal D_a(\la)=(\la_1+\th, \dots,\la_{j-1}+\th, \la_{j+1}, \la_{j+2},
\dots).
$$
Note that
$$
|\Cal D_a(\la)|=|\la|-a-\th.
$$

More generally, let $A=\{a_1,\dots,a_k\}$ be a $k$--tuple of integral
points such that the pairwise distances between them are at least
$\th$.  Given a diagram $\la$ such that $\LL(\la)$ contains $A$ we
define a new diagram $\Cal D_A(\la)$ as follows: $\LL(\Cal D_A(\la))$ is
obtained from $\LL(\la)$ by removing $A$ and shifting the remaining points by
$k\th$.
Clearly,
$$
\Cal D_A=\Cal D_{a_k+(k-1)\th}\circ\dots\circ
\Cal D_{a_2+\th}\circ\Cal D_{a_1}\,.
$$
It follows, in particular, that
$$
|\Cal D_A(\la)|=|\la|-a_1-\dots-a_k-\frac{k(k+1)}2\th.
$$

\proclaim{Proposition 3.2} Fix a $k$-point subset $A$ of $\Z$. A Young diagram
$\mu$ can be represented
as $\Cal D_A(\la)$ for a Young diagram $\la$ if and only if $\LL(\mu)$ does not
intersect the set
$$
\bigcup_{j=1}^k [a_j+(k-1)\th+1,\, a_j+(k+1)\th-1].
$$
\endproclaim
\demo{Proof} Evident.\qed\enddemo

For any Young diagram $\la$ we introduce a rational function
$$
E_\th^*(\la;u)=\prod_{i=1}^{\infty}\frac{u+\la_i-i\th+\th}{u-i\th+\th}
=\prod_{i=1}^\infty\frac{u+l_i+\th}{u-i\th+\th}\,.
$$
Both these products are, in fact, finite, because the $i$th factor turns into 1
as soon as $i>\ell(\la)$. This function has no poles in $\{u\in\C:\Re u<0\}$. As
we will see later, $E_\th^*(\la;u)$ is a discrete counterpart of the function
$E_\th(\om;u)$ introduced in \S2.

We also define
$$
\tE(\la;u)=\frac{E_\th^*(\la;u)}{\Ga(-u/\th)}\,.
\tag 3.1
$$

\proclaim{Proposition 3.3}
For any Young diagram $\la$, $\tE(\la;u)$ is an entire
function in $u$. It has simple zeros at the points
$u=-l_i-\th=-\la_i+i\th-\th$, where $i=1,2,\dots$. Moreover, these
are the only zeros of $\tE(\la;u)$.
\endproclaim
\demo{Proof} Fix $\la$ and let $r$ be a large enough integer. We have
$$
\gather
\tE(\la;u)=\frac1{\Ga(-u/\th)}\,
\prod_{i=1}^r\frac{u+l_i+\th}{u-i\th+\th}\\
=\frac1{\Ga(-u/\th)}\,
\prod_{i=1}^r\frac{-u/\th-l_i/\th-1}{-u/\th+i-1}\\
=\frac1{\Ga(-u/\th+r)}\prod_{i=1}^r(-u/\th-l_i/\th-1)
\endgather
$$

This expression is clearly an entire function in $u$. Restrict $u$ to a left
half--plane of the form $\Re u\le c$ where $c\gg0$. The above argument with
large enough $r$ shows that the factor $\frac1{\Ga(-u/\th+r)}$ does not vanish
in that half--plane. Thus, the only zeros come from the product. But these are
simple zeros at $u=-l_i-\th$. \qed
\enddemo

For any function $F$ on the set $\Y$ of all Young diagrams we denote by
$\langle F\rangle_{z,z',\th;\xi}$ the average value of $F$ with respect to $\wt
M_{z,z',\th;\xi}$:
$$
\langle F\rangle_{z,z',\th;\xi}=\sum_{\la\in\Y}F(\la)
\wt M_{z,z',\th;\xi}(\la).
$$

The next statement expresses the correlation functions of the mixed measures
$\wt M_{z,z',\th;\xi}$ through the averages of products of
$\tE$ with appropriate arguments.

\proclaim{Theorem 3.4}Let $A=\{a_1,\dots,a_k\}$ be a $k$-point subset of $\Z$.
We have
$$
\wt M_{z,z',\th;\xi}\left(\{\la\in\Y\mid \LL(\la)\supset A\}\right)=
C\left\langle\prod_{j=1}^k\prod_{\sigma=0}^{\th-1}
\tE(\,\cdot\,;u^+_{j,\sigma})\tE(\,\cdot\,;u^-_{j,\sigma})
\right\rangle_{z-k\th,z'-k\th,\th;\xi}
$$
where the prefactor $C$ is given by
$$
\gathered
C=(2\pi)^{k(\th-1)}(\Gamma(\th))^k\th^{-2(a_1+\dots+a_k)-\th k(2k+1)}(1-
\xi)^{k(z+z')-k^2\th}\, \xi^{a_1+\dots+a_k+k(k+1)\th/2}\\ \times
\prod_{j=1}^k\frac{\Ga(z+a_j+\th)\Ga(z'+a_j+\th)}
{\Ga(z-j\th+\th)\Ga(z'-j\th+\th)}
\,\cdot
\prod_{1\le j<j'\le k}\prod_{\sigma=0}^{\th-1}((a_j-a_{j'})^2-\sigma^2).
\endgathered
$$
and
$$
u^\pm_{j,\sigma}=-a_j\pm\sigma-(k+1)\th, \qquad j=1,\dots,k, \quad
\sigma=0,1,\dots,\th-1.
$$
\endproclaim
\demo{Proof}
The claim is equivalent to
$$
\sum_{\la:\, \LL(\la)\supset A} \wt M_{z,z',\th;\xi}(\la)=
C\sum_{\mu\in\Y}\prod_{j=1}^k
\prod_{\sigma=0}^{\th-1} \tE(\mu;u^+_{j,\sigma})\tE(\mu;u^-_{j,\sigma})\,\cdot\,
\wt M_{z-k\th,z'-k\th,\th;\xi}(\mu).
$$

If $\LL(\mu)$ intersects
$$
\bigcup_{j=1}^k [a_j+(k-1)\th+1,\,a_j+(k+1)\th-1]
$$ then
one of the factors $\tE(\mu;u^\pm_{j,\sigma})$ vanishes by Proposition 3.3.
Hence, we may consider only those $\mu$ which are of the form $\bla:=\Cal
D_A(\la)$.

Thus, it suffices to prove that for any $\la$ such that $\LL(\la)$ contains
$A$,
$$
{\wt M_{z,z',\th;\xi}(\la)}
=C\prod_{j=1}^k\prod_{\sigma=0}^{\th-1}
\tE(\bla;u^+_{j,\sigma})\tE(\bla;u^-_{j,\sigma})\cdot {\wt M_{z-k\th,z'-
k\th,\th;\xi}(\bla)}.
$$

By the definition of $\wt M_{z,z',\th;\xi}$, we have
$$
\wt M_{z,z',\th;\xi}(\la)=\underbrace{(1-
\xi)^{zz'/\th}\xi^{|\la|}}_{\text{(1)}}\,\cdot\,
\underbrace{(z)_{\la,\th}(z')_{\la,\th}}_{\text{(2)}}\,\cdot\,
\underbrace{\frac1{H(\la;\th)H'(\la;\th)}}_{\text{(3)}}\,.
$$
Similarly,
$$
\multline \wt
M_{z-k\th,z'-k\th,\th;\xi}(\bla)\\=\underbrace{(1-\xi)^{(z-k\th)(z'-
k\th)/\th}\xi^{|\bla|}}_{\text{(1)}}\,\cdot\,
\underbrace{(z-k\th)_{\bla,\th}(z'-k\th)_{\bla,\th}}_{\text{(2)}}\,\cdot\,
\underbrace{\frac1{H(\bla;\th)H'(\bla;\th)}}_{\text{(3)}}\,.
\endmultline
$$

The ratio of the first factors is
$$
\frac{(1-\xi)^{zz'/\th}\,\xi^{|\la|}}
{(1-\xi)^{(z-k\th)(z'-k\th)/\th}\,\xi^{|\bla|}}
=(1-\xi)^{k(z+z')-k^2\th}\, \xi^{a_1+\dots+a_k+k(k+1)\th/2}\,.
$$
We used the fact that $|\bla|=|\la|-(a_1+\dots+a_k)-\frac{k(k+1)}2\th$ mentioned
above.

To handle the second factors, let us rewrite these factors
in terms of $\LL(\la)$, $\LL(\bla)$. Denote
$$
\LL(\la)=\{l_1,l_2,\dots\}, \qquad
\LL(\bla)=\{\bl_1,\bl_{2}, \dots\}.
$$

With this notation, for any integral $r$ large enough we can write
$$
\gathered
(z)_{\la,\th}(z')_{\la,\th}=\prod_{i=1}^r
\frac{\Gamma(z+l_i+\th)}{\Ga(z-i\th+\th)}\,
\frac{\Gamma(z'+l_i+\th)}{\Ga(z'-i\th+\th)}\,,\\
(z-k\th)_{\bla,\th}(z'-k\th)_{\bla,\th}=\prod_{i=1}^{r-k}
\frac{\Gamma(z-k\th+\bar l_i+\th)}{\Ga(z-k\th-i\th+\th)}\,
\frac{\Gamma(z'-k\th+l_i+\th)}{\Ga(z'-k\th-i\th+\th)}\,.
\endgathered
$$
Observe that for a large integer $r$ the numbers $\bl_1,\dots,\bl_{r-k}$ are
obtained
from the numbers $l_1,\dots,l_r$ by removing $a_1,\dots,a_k$ and adding
$k\th$ to each of the $r-k$ remaining numbers. This implies that
$$
(z)_{\la,\th}(z')_{\la,\th}=\prod_{j=1}^k\frac{\Ga(z+a_j+\th)\Ga(z'+a_j+\th)}
{\Ga(z-j\th+\th)\Ga(z'-j\th+\th)}\cdot
(z-k\th)_{\bla,\th}(z'-k\th)_{\bla,\th}.
$$

The ratio of the third factors is computed in
\proclaim{Lemma 3.5} For any large enough integer $r$, we have
$$
\gathered
\frac{H(\bla;\th)H'(\bla;\th)}{H(\la;\th)H'(\la;\th)}
=(\Ga(\th))^k
\prod_{1\le j<j'\le k}\prod_{\sigma=0}^{\th-1}((a_j-a_{j'})^2-\sigma^2)\\
\times
\prod_{i=1}^{r-k}\prod_{j=1}^k\prod_{\sigma=0}^{\th-1}
((\bl_i-a_j-k\th)^2-\sigma^2)
\cdot\prod_{j=1}^k\frac{1}{\Ga(a_j+r\th+1)\Ga(a_j+r\th+\th)}\,.
\endgathered
$$
\endproclaim
\demo{Proof}
$$
\gather
H(\la;\th)=\prod_{1\le i<j\le r}\frac{((j-i)\th+1-\th)_{\la_i-\la_j}}
{((j-i)\th+1)_{\la_i-\la_j}}\,\cdot\,
\prod_{i=1}^r((r-i)\th+1)_{\la_i}\\
=\prod_{1\le i<j\le r} \frac{\Ga(l_i-l_j+1-\th)}{\Ga(l_i-l_j+1)}\cdot\prod_{1\le
i<j\le r}\frac{\Ga((j-i)\th+1)}
{\Ga((j-i)\th+1-\th)}\cdot\prod_{i=1}^r\frac{\Ga(l_i+r\th+1)}{\Ga((r-
i)\th+1)}\,,
\endgather
$$
The first product is equal to
$$
\prod_{1\le i<j\le r}\prod_{\sigma=0}^{\th-1}\frac 1{l_i-l_j-\sigma}\,.
$$
The second product is equal to
$$
\prod_{1\le i<j\le r}\frac{\Ga((j-i)\th+1)}
{\Ga((j-i-1)\th+1)}=\prod_{i=1}^r\Ga((r-i)\th+1).
$$
Hence, we obtain
$$
H(\la;\th)=\prod_{1\le i<j\le r}\prod_{\sigma=0}^{\th-1}\frac 1{l_i-l_j-
\sigma}\cdot\prod_{i=1}^r\Ga(l_i+r\th+1)\,.
$$
Likewise,
$$
\gather
H'(\la;\th)=\prod_{1\le i<j\le r}\frac{((j-i)\th)_{\la_i-\la_j}}
{((j-i)\th+\th)_{\la_i-\la_j}}\,\cdot\,
\prod_{i=1}^r((r-i)\th+\th)_{\la_i}\\
=\prod_{1\le i<j\le r} \frac{\Ga(l_i-l_j)}{\Ga(l_i-l_j+\th)}\cdot \prod_{1\le
i<j\le r}\frac{\Ga((j-i)\th+\th)}
{\Ga((j-i)\th)}\cdot
\prod_{i=1}^r\frac{\Ga(l_i+r\th+\th)}{\Ga((r-i)\th+\th)}\\
=\prod_{1\le i<j\le r}\prod_{\sigma=0}^{\th-1}\frac 1{l_i-
l_j+\sigma}\cdot\prod_{i=1}^r\frac{\Ga(l_i+r\th+\th)}{\Ga(\th)}\,.
\endgather
$$
Therefore,
$$
H(\la;\th)H'(\la;\th)=\prod_{1\le i<j\le r}\prod_{\sigma=0}^{\th-1}\frac 1{(l_i-
l_j)^2-\sigma^2}\cdot\prod_{i=1}^r
\frac{\Ga(l_i+r\th+1)\Ga(l_i+r\th+\th)}{\Ga(\th)}\,.
$$
Similarly, for $\bla$ we get
$$
\multline
H(\bla;\th)H'(\bla;\th)\\=\prod_{1\le i<j\le r-k}\prod_{\sigma=0}^{\th-1}\frac
1{(\bl_i-\bl_j)^2-\sigma^2}\cdot\prod_{i=1}^{r-k}
\frac{\Ga(\bl_i+(r-k)\th+1)\Ga(\bl_i+(r-k)\th+\th)}{\Ga(\th)}\,.
\endmultline
$$
Using the observation made before the statement of Lemma 3.5,
we readily obtain the needed result.\qed
\enddemo
\proclaim{Lemma 3.6} For any large enough integer $r$, we have
$$
\gather
\prod_{j=1}^k\prod_{\sigma=0}^{\th-1}
\tE(\bla;u^+_{j,\sigma})\tE(\bla;u^-_{j,\sigma})=(2\pi)^{k(1-
\th)}\th^{2(a_1+\dots+a_k)+\th k(2k+1)}\\
\times
\prod_{i=1}^{r-k}\prod_{j=1}^k\prod_{\sigma=0}^{\th-1}
((\bl_i-a_j-k\th)^2-\sigma^2)
\cdot\prod_{j=1}^k\frac{1}{\Ga(a_j+r\th+1)\Ga(a_j+r\th+\th)}
\endgather
$$
\endproclaim
\demo{Proof} We have, cf. the proof of Proposition 3.3,
$$
\tE(\bla;u)=\frac{\prod\limits_{i=1}^{r-k}(u+\bl_i+\th)}{(-\th)^{r-k}\Ga(-
u/\th+r-k)}\,.
$$
Note that
$$
u_{j,\sigma}^\pm+\bl_i+\th=-a_j+\bl_i-k\th\pm\sigma,\qquad-
\frac{u_{j,\sigma}^\pm}{\th}+r-k=\frac{a_j\mp\sigma}\th+r+1.
$$
Hence,
$$
%\multline
\prod_{j=1}^k\prod_{\sigma=0}^{\th-1}
\tE(\bla;u^+_{j,\sigma})\tE(\bla;u^-_{j,\sigma})=\frac{\th^{-2\th k(r-
k)}\prod\limits_{i=1}^{r-k}\prod\limits_{j=1}^k\prod\limits_{\sigma=0}^{\th-1}
((\bl_i-a_j-k\th)^2-\sigma^2)}{\prod\limits_{j=1}^k\prod\limits_{\sigma=0}^{\th-
1}\Ga\left(\dfrac{a_j-
\sigma}\th+r+1\right)\Ga\left(\dfrac{a_j+\sigma}\th+r+1\right)}\,.
$$
Applying the multiplication formula for the gamma-function
$$
\prod_{\sigma=0}^{\th-1}\Ga\left(x+\frac \sigma\th\right)
=(2\pi)^{\frac{\th-1}2}\th^{\frac 12-\th x}\Ga(\th x)
\tag 3.2
$$
in the denominator, we obtain the result.\qed
\enddemo

Putting all these computations together, we arrive at the formula of Theorem
3.4.\qed
\enddemo

To conclude this section, we restate Theorem 3.4 in terms of averages of
$E_\th^*(\,\cdot\,;u)$ rather than $\tE(\,\cdot\,;u)$. Because of that,
we have to restrict ourselves to subsets $A$ of $\Z_{\ge 0}$, not of $\Z$, but
the new formulation will be more convenient for the limit transition in \S4.

\proclaim{Corollary 3.7}
Let $A=\{a_1,\dots,a_k\}$ be a $k$-point subset of $\Z_{\ge 0}$. We have
$$
\wt M_{z,z',\th;\xi}\left(\{\la\in\Y\mid \LL(\la)\supset A\}\right)=
C'\left\langle\prod_{j=1}^k\prod_{\sigma=0}^{\th-1}
E_\th^*(\,\cdot\,;u^+_{j,\sigma})E_\th^*(\,\cdot\,;u^-_{j,\sigma})
\right\rangle_{z-k\th,z'-k\th,\th;\xi}
$$
where the prefactor $C'$ is given by
$$
\gathered
C'=(1-\xi)^{k(z+z')-k^2\th}\, \xi^{a_1+\dots+a_k+k(k+1)\th/2}\prod_{j=1}^k\frac
{\Ga(\th)}{\Ga(a_j+k\th+1)\Ga(a_j+k\th+\th)}\\ \times
\prod_{j=1}^k\frac{\Ga(z+a_j+\th)\Ga(z'+a_j+\th)}
{\Ga(z-j\th+\th)\Ga(z'-j\th+\th)}
\,\cdot
\prod_{1\le j<j'\le k}\prod_{\sigma=0}^{\th-1}((a_j-a_{j'})^2-\sigma^2).
\endgathered
$$
and
$$
u^\pm_{j,\sigma}=-a_j\pm\sigma-(k+1)\th, \qquad j=1,\dots,k, \quad
\sigma=0,1,\dots,\th-1.
$$
\endproclaim
\demo{Proof} First of all, recall that $E_\th^*(\,\cdot\,;u)$ is a meromorphic
function in $u$ which has no poles in $\{u\in\C:\Re u<0\}$.
Because of that, the product of $E_\th^*$ above makes sense if all $a_i$ are
nonnegative. Indeed, then $u_{j,\sigma}^\pm<0$ for all $j,s$.

By \tht{3.1}, we have
$$
\prod_{j=1}^k\prod_{\sigma=0}^{\th-1}
\tE(\,\cdot\,;u^+_{j,\sigma})\tE(\,\cdot\,;u^-_{j,\sigma})=
\frac{
\prod_{j=1}^k\prod_{\sigma=0}^{\th-1}
E_\th^*(\,\cdot\,;u^+_{j,\sigma})E_\th^*(\,\cdot\,;u^-
_{j,\sigma})}{\prod_{j=1}^k\prod_{\sigma=0}^{\th-1}
\Ga(-u^+_{j,\sigma}/\th)\Ga(-u^-_{j,\sigma}/\th)}\,.
$$
Applying the multiplication formula for the gamma-function \tht{3.2}, we obtain
$$
\multline
\prod_{j=1}^k\prod_{\sigma=0}^{\th-1}
\Ga(-u^+_{j,\sigma}/\th)\Ga(-u^-_{j,\sigma}/\th)\\=(2\pi)^{k(\th-1)}\th^{-
2(a_1+\dots+a_k)-\th k(2k+1)}\prod_{j=1}^k\Ga(a_j+k\th+1)\Ga(a_j+k\th+\th).
\endmultline
$$
Thus, Theorem 3.4 implies the needed claim with $C'$ equal to $C$ divided by the
expression above. \qed
\enddemo

\head 4. Convergence of correlation functions
\endhead

The goal of this section is to prove that the lattice correlation functions
$$
M_{z,z',\th}^{(n)}\left(\{\la\in\Y_n\mid \LL(\la)\supset
\{x_1,\dots,x_k\}\}\right), \,
\wt M_{z,z',\th;\xi}\left(\{\la\in\Y\mid \LL(\la)\supset
\{x_1,\dots,x_k\}\}\right)
$$
converge, in the corresponding scaling limits as $n\to\infty$ or $\xi\nearrow
1$, to the correlation functions
$$
\rho_k(y_1,\dots,y_k),\qquad \wt \rho_k(y_1,\dots,y_k)
$$
defined in the end of \S1.

For the random Young diagram $\la\in\Y_n$ distributed according to
$M_{z,z',\th}^{(n)}$ introduce the random variables
$$
\al_i^{(n)}=\cases \dfrac{l_i-i\th}n,&l_i-i\th>0,\\
0,&\text{otherwise},
\endcases
$$
where $\{l_1,l_2,\dots\}=\LL(\la)$. These $\al_i^{(n)}$ are different from
those introduced in Remark 1.7 by $O(1/n)$. Thus, by Remark 1.7, we still have
for any positive integer $m$ the convergence
$$
\{\al_1^{(n)},\dots,\al_m^{(n)}\}\overset d\to\longrightarrow
\{\al_1,\dots,\al_m\}.
\tag 4.1
$$

Let
$r_k^{(n)}$ denote the $k$th correlation measure for
$\left\{\al_i^{(n)}\right\}_{i=1}^\infty$. Formally, for any compactly supported
continuous function $F$ on $(\R_{>0})^k$,
$$
\langle F,r_k^{(n)}\rangle =\E_n\left(\sum_{\Sb
i_1,i_2,\dots,i_k\\\text{pairwise distinct}\endSb}
F(\al_{i_1}^{(n)},\dots,\al_{i_k}^{(n)})\right),
\tag 4.2
$$
where $\E_n$ denotes the expectation with respect to
$M_{z,z',\th}^{(n)}$.

Recall that the $k$th correlation measure for $\{\al_i\}$ was defined in a
similar way in \S1:
$$
\langle F,\rho_k\rangle=\E\left(\sum_{\Sb i_1,i_2,\dots,i_k\\\text{pairwise
distinct}\endSb}
F(\al_{i_1},\dots,\al_{i_k})\right),
\tag 4.3
$$
where $\E$ denotes the expectation with respect to $\M_{z,z',\th}$.
\proclaim{Proposition 4.1} For any $k=1,2,\dots$, and any compactly supported
continuous function $F$ on $(\R_{>0})^k$, we have
$$
\langle F,r_k^{(n)}\rangle\longrightarrow
\langle F,\rho_k\rangle ,\qquad n\to\infty.
$$
\endproclaim
\demo{Proof} We rely on the convergence of the finite-dimensional distributions
\tht{4.1} and the fact that
$$
\gathered
\al_1^{(n)}\ge \al_2^{(n)}\ge\dots\ge 0,\qquad\sum_{i=1}^\infty \al_i^{(n)}\le
1,\\
\al_1\ge\al_2\ge\dots\ge 0,\qquad\sum_{i=1}^\infty\al_i\le 1.
\endgathered
\tag 4.4
$$
These inequalities imply that
$$
\al_{m+1}^{(n)}< 1/m,\quad \al_{m+1}< 1/m,\qquad m=1,2,\dots\,,
\tag 4.5
$$
cf. \tht{1.2}.
Fix $m$ so large that $\operatorname{supp} F\subset (\R_{\ge 1/m})^k$.
Then the summands in \tht{4.2} and \tht{4.3} involving indices $i_l>m$ vanish.
Thus, only finitely many summands remain, and the statement follows from
\tht{4.1}.\qed
\enddemo

We proceed to the mixed measures $\wt M_{z,z',\th;\xi}$.
For the random Young diagram $\la\in\Y$ distributed according to
$M_{z,z',\th;\xi}$ introduce the random variables
$$
\al_{i,\xi}=\cases (1-\xi)(l_i-i\th),&l_i-i\th>0,\\
0,&\text{otherwise},
\endcases
$$
where $\{l_1,l_2,\dots\}=\LL(\la)$ as above.
We define the mixed correlation measures $\wt r_k^{(\xi)}$, $k=1,2,\dots$, by
$$
\langle F,\wt r_k^{(\xi)}\rangle =\E_\xi\left(\sum_{\Sb
i_1,i_2,\dots,i_k\\\text{pairwise distinct}\endSb}
F(\al_{i_1,\xi},\dots,\al_{i_k,\xi})\right),
$$
where $\E_\xi$ denotes the expectation with respect to
$M_{z,z',\th;\xi}$. These are essentially the same objects as in Theorem 3.4,
with the lattice $\Z$ being scaled by $(1-\xi)$.

Recall that the lifted correlation functions (measures) $\wt \rho_k$ were
defined in the end of \S1.

\proclaim{Proposition 4.2}
For any $k=1,2,\dots$, and any compactly supported continuous function $F$ on
$(\R_{>0})^k$, we have
$$
\langle F,\wt r_k^{(\xi)}\rangle\longrightarrow
\langle F,\wt\rho_k\rangle ,\qquad n\to\infty.
$$
\endproclaim
\demo{Proof} Let
$$
\ga_t=\frac{s^{t-1}}{\Ga(t)}e^{-s}ds
$$
be the gamma-distribution on $\R_{>0}$ with the parameter $t=zz'/\th$, and let
$$
\ga_{t,\xi}=(1-\xi)^t\sum_{n=0}^\infty\frac{(t)_n}{n!}\,\xi^n\delta_{n(1-\xi)}
$$
be a scaled version of the negative binomial distribution. Here $\delta_x$
stands for the Dirac measure at $x$.
The similarity of notation is justified by the following statement.

\proclaim{Lemma 4.3}{\rm (i)} The distribution $\gamma_{t,\xi}$ weakly
converges to $\gamma_t$ as $\xi\nearrow1$.

{\rm (ii)} All moments of the distribution $\gamma_{t,\xi}$  converge to the
respective moments of $\gamma_t$ as $\xi\nearrow1$.
\endproclaim

\demo{Proof of Lemma 4.3} (i) For any $s>0$ we define $n(s,\xi)=[s/(1-\xi)]$.
Since both $\gamma_{t,\xi}$ and $\gamma_t$ are {\it probability} measures, it
suffices to show that
$$
(1-\xi)^t\frac{(t)_{n}}{n!}\,\xi^{n}\cdot(1-\xi)^{-1}\longrightarrow
\frac{s^{t-1}}{\Ga(t)}e^{-s},\qquad n=n(s,\xi),\quad \xi\nearrow 1,
$$
for any $s>0$.
Indeed, we have, with $n=n(s,\xi)$ and $\xi\nearrow 1$,
$$
\multline
(1-\xi)^{t-1}\frac{(t)_{n}}{n!}\,\xi^{n}=
\frac{(1-\xi)^{t-1}}{\Ga(t)}\frac{\Ga(n+t)}{\Ga(n+1)}\,(1-(1-\xi))^n\\ \sim
\frac{(1-\xi)^{t-1}n^{t-1}e^{-s}}{\Ga(t)}\sim \frac{s^{t-1}e^{-s}}{\Ga(t)}\,.
\endmultline
$$
(ii) We have to prove that for any $m=1,2,\dots$,
$$
\lim_{\xi\nearrow 1}\left(
(1-\xi)^t\sum_{n=0}^\infty\frac{(t)_n}{n!}\,\xi^n(n(1-\xi))^m\right)=
\int_{0}^\infty \frac{s^{t-1}}{\Ga(t)}\,s^me^{-s}ds=(t)_m.
$$
Note that
$$
(n(1-\xi))^m=(1-\xi)^mn(n-1)\cdots(n-m+1)\cdot(1+O(1-\xi))
$$
uniformly in $n=0,1,\dots$\,. Thus, it suffices to show that
$$
\lim_{\xi\nearrow 1}\left(
(1-\xi)^{t+m}\sum_{n=0}^\infty\frac{(t)_n}{n!}\,\xi^n n(n-1)\cdots (n-
m+1)\right)=(t)_m.
$$
But the sum in the left--hand side is easily computed:
$$
\sum_{n=0}^\infty\frac{(t)_n}{n!}\,\xi^n n(n-1)\cdots (n-m+1)
=(t)_m\xi^m\sum_{l=0}^\infty \frac{(t+m)_l}{l!}\,\xi^l=(t)_m\xi^m(1-\xi)^{-t-m}.
$$
The needed limit relation immediately follows.\qed
\enddemo
Let us return to the proof of Proposition 4.2. We have
$$
\langle F,\wt r_k^{(\xi)}\rangle=
\int_{0}^\infty \E_{n(s,\xi)}\left(\sum_{\Sb i_1,\dots,i_k\\ \text{pairwise
distinct}\endSb}F\left(s\cdot \al_{i_1}^{(n(s,\xi))},\dots,s\cdot
\al_{i_k}^{(n(s,\xi))}\right)\right)\ga_{t,\xi}(ds).
$$
Note that for $s\in\operatorname{supp}(\gamma_{t,\xi})$,
$n(s,\xi)=[s/(1-\xi)]=s/(1-\xi)$.

Similarly,
$$
\langle F,\wt \rho_k\rangle=
\int_{0}^\infty \E\left(\sum_{\Sb i_1,\dots,i_k\\ \text{pairwise
distinct}\endSb}F(s\cdot\al_{i_1},\dots,s\cdot\al_{i_k})\right)\ga_{t}(ds).
$$
Fix $\epsilon>0$ so small that $\operatorname{supp}F\subset (\R_{\ge
\epsilon})^k$. Since $\al_i^{(n)}\le 1$, $\al_i\le 1$, both integrals remain
intact if we replace the lower limit of integration by $\epsilon$.

\proclaim{Lemma 4.4}
For any $S>\epsilon$, we have
$$
\multline
\lim_{\xi\nearrow 1}\int_{\epsilon}^S \E_{n(s,\xi)}\left(\sum_{\Sb i_1,\dots,i_k\\
\text{pairwise distinct}\endSb}F\left(s\cdot
\al_{i_1}^{(n(s,\xi))},\dots,s\cdot
\al_{i_k}^{(n(s,\xi))}\right)\right)\ga_{t,\xi}(ds)\\= \int_{\epsilon}^S
\E\left(\sum_{\Sb i_1,\dots,i_k\\ \text{pairwise
distinct}\endSb}F(s\cdot\al_{i_1},\dots,s\cdot\al_{i_k})\right)\ga_{t}(ds).
\endmultline
$$
\endproclaim
\demo{Proof of Lemma 4.4}
By the argument in the proof of Proposition 4.1, the sums above are actually
finite, and it suffices to prove the limit relation for any fixed indices
$i_1,\dots,i_k$, that is, we will show that
$$
\multline
\lim_{\xi\nearrow 1}\int_{\epsilon}^S \E_{n(s,\xi)}\left(F\left(s\cdot
\al_{i_1}^{(n(s,\xi))},\dots,s\cdot
\al_{i_k}^{(n(s,\xi))}\right)\right)\ga_{t,\xi}(ds)\\=
\int_{\epsilon}^S
\E\left(F(s\cdot\al_{i_1},\dots,s\cdot\al_{i_k})\right)\ga_{t}(ds).
\endmultline
$$
It is convenient to denote $F_s(x_1,\dots,x_k)=F(s\cdot x_1,\dots,s\cdot x_k)$.
Since $F$ is compactly supported, the map $s\mapsto F_s$ is continuous on
$[\epsilon,S]$ with respect to the sup-norm in the Banach space of continuous
functions. Therefore, $\{F_s,s\in[\epsilon,S]\}$ is a compact set. Hence, by
\tht{4.1}, $\E_n(F_s(\al_{i_1}^{(n)},\dots,\al_{i_k}^{(n)}))$ is close to
$\E(F_s(\al_{i_1},\dots,\al_{i_k}))$ for large $n$ uniformly in $s\in[\epsilon,
S]$.

Since the variable of integration $s$ is bounded away from zero, $n(s,\xi)$ is
uniformly large as $\xi\nearrow 1$. Thus, it suffices to show that
$$
\lim_{\xi\nearrow 1}\int_{\epsilon}^S
\E\left(F(s\cdot\al_{i_1},\dots,s\cdot\al_{i_k})\right)\ga_{t,\xi}(ds)=
\int_{\epsilon}^S
\E\left(F(s\cdot\al_{i_1},\dots,s\cdot\al_{i_k})\right)\ga_{t}(ds).
$$
Since the integrand is continuous in $s$, the convergence follows from Lemma
4.3(i).
\qed
\enddemo

To complete the proof of the Proposition 4.2, it remains to prove that
$$
\int_{S}^\infty \E_{n(s,\xi)}\left(\sum_{\Sb i_1,\dots,i_k\\ \text{pairwise
distinct}\endSb}F\left(s\cdot \al_{i_1}^{(n(s,\xi))},\dots,s\cdot
\al_{i_k}^{(n(s,\xi))}\right)\right)\ga_{t,\xi}(ds)\longrightarrow 0
$$
as $S\to\infty$, uniformly in $\xi$.

Observe that for any fixed $s$ the number of terms in the sum above is $O(s^k)$
independently of $\xi$. Indeed, recall that $\al_{m+1}^{(n)}< 1/m$, see
\tht{4.5}. On the other hand, we must have $s\al_{i_l}^{(n)}\ge \epsilon$ in
order for the corresponding term not to vanish. Thus, we are only allowed to
take $i_l\le s/\epsilon$.

Thus, the absolute value of the integral is bounded by
$$
\operatorname{const}\cdot\int_{S}^\infty s^k \ga_{t,\xi}(ds),
$$
and the result readily follows from Lemma 4.3(ii). The proof of Proposition 4.2
is complete.
\qed
\enddemo

\head 5. Limit correlation functions
\endhead

The goal of this section is to derive hypergeometric-type formulas for the limit
correlation functions.

Our first step is to define the limit of the right--hand side of the formula in
Corollary 3.7.

We will use the notation (the function $E_\th(\om;u)$ was introduced in \S2)
$$
\gather
E^*(\la;u)=E_{\th}^*(\la;u)\bigl|_{\th=1}=\prod_{i=1}^\infty\frac{u-\la_i-
i+1}{u-i+1}\,
,\\
E(\om;u)=E(\om;u)\bigl|_{\th=1}=
e^{\ga/u}\frac{\prod_{i=1}^\infty(1+\al_i/u)}{\prod_{i=1}^\infty(1-\be_i/u)}\,,
\endgather
$$
and
$$
\gather
\la_\th=(\underbrace{\la_1,\dots,\la_1}_\th\,,\underbrace{\la_2,\dots,\la_2}_\th
\,,\dots),\quad \la\in\Y,\\
\al_\th=(\underbrace{\al_1,\dots,\al_1}_\th\,,\underbrace{\al_2,\dots,\al_2}_\th
\,,\dots),\quad \th\be=(\th\be_1,\th\be_2,\dots),
\quad\om_\th=(\al_\th,\th\be,\th\delta).
\endgather
$$

Recall that in \S1 we defined the modified Frobenius coordinates
$\{a_i(\la);b_i(\la)\}$ of a Young diagram $\la$. Set
$$
\iota(\la)=(a_1(\la),a_2(\la),\dots;b_1(\la),b_2(\la),\dots;|\la|)\in\wOm.
$$

\proclaim{Proposition 5.1}
$$
\gather
E_\th^*(\la;u)E_\th^*(\la;u-1)\cdots E_\th^*(\la;u-\th+1)=E^*(\la_\th;u),\qquad
\la\in\Y,\\
\left(E_\th(\om;u)\right)^\th= E(\om_\th;u),\qquad \om\in\wOm,\\
E^*(\la;u)=E(\iota(\la);u+\tfrac 12),\qquad \la\in\Y.
\endgather
$$
\endproclaim
\demo{Proof} The first relation readily follows from the definition of
$\la_\th$. The second relation is evident. The third relation is also not hard
to prove, see, e.g., \cite{ORV}.\qed
\enddemo

The third relation shows that $E^*$ and $E$ are essentially the same, if the
Young diagrams are viewed as points of $\wOm$ via the embedding $\iota$.

The next statement computes the limit of the expectation in right--hand side of
Corollary 3.7. (To simplify the notation, we temporarily ignore the shift of the
parameters $z,z'$ in Corollary 3.7.)

\proclaim{Proposition 5.2} For any $k=1,2,\dots$, and sufficiently large
$x_1,\dots,x_k>0$, if $a_i=a_i(\xi)$, $i=1,\dots,k$, are such that
$a_i(1-\xi)\to x_i$ as $\xi\nearrow 1$, then
$$
\gathered
\lim_{\xi\nearrow 1}\left\langle\prod_{j=1}^k\prod_{\sigma=0}^{\th-1}
E_\th^*(\,\cdot\,;u^+_{j,\sigma})E_\th^*(\,\cdot\,;u^-_{j,\sigma})
\right\rangle_{z,z',\th;\xi}=
\int\limits_{\om\in\wOm} \prod_{j=1}^k (E_\th(\om;-x_j))^{2\th}\,
\wt\M_{z,z',\th}(d\om),
\endgathered
\tag 5.1
$$
where
$$
u^\pm_{j,\sigma}=-a_j\pm \sigma-(k+1)\th, \qquad j=1,\dots,k, \quad
\sigma=0,1,\dots,\th-1.
$$
\endproclaim

We will need the following simple lemma. Recall that in \S1 we introduced a
metric on $\wOm$ denoted by $\dist(\,\cdot\,\,,\,\cdot\,)$.
\proclaim{Lemma 5.3}
{\rm (i)} For any $\om\in\wOm$ and $u<0$, we have
$$
|E(\om;u)|\le e^{\delta(\om)/|u|}
$$
where, as above, $\delta(\om)$ denotes the $\delta$-coordinate of $\om$.

{\rm (ii)} Assume that $\dist(\om',\om'')\to 0$ and $u'-u''\to 0$. Then
$$
E(\om';u')-E(\om'';u'')\to 0
$$
uniformly on any set of the form
$\{\om\in\wOm:\delta(\om)\le \const_1\}\times\{u\le\const_2<0\}$.
\endproclaim
\demo{Proof} (i) Without loss of generality we may assume that $\gamma(\om)=0$,
because this condition defines a dense subset of $\wOm$. By the 0-homogeneity of
$E(\om;u)$, we may also assume that $u=-1$.
Let $m=m(\om)$ be the number of $\al_i=\al_i(\om)$ which are greater than 1.
Then
$$
|E(\om;-1)|=\left|\frac{\prod_{i=1}^\infty(1-\alpha_i)}
{\prod_{i=1}^\infty(1+\beta_i)}\right|\le \prod_{i=1}^m \alpha_i\le
\left(\frac{\sum_{i=1}^m \al_i}m\right)^m\le \frac{\delta^m}{m!}\le
e^\delta.\qed
$$

(ii) By homogeneity, we have
$$
E(\om';u')=E(\om'/|u'|;-1),\qquad
E(\om'';u'')=E(\om''/|u''|;-1).
$$
The statement now follows from the uniform continuity of the function
$E(\om;-1)$ on the compact set $\{\om\in\wOm:\delta(\om)\le\const\}$.\qed
\enddemo

\example{Remark 5.4}
Even though the estimate of (i) above seems rather coarse, it cannot
be substantially improved: one can show that
$\sup\{ E(\om;-1)\mid\delta(\om)=\Delta\}$ grows at least as $e^{\const
\cdot\Delta}$ as $\Delta\to\infty$.
As we will see below, this is the reason why we have to assume that $x_i$'s are
large in the proof of Proposition 5.2.
\endexample

\demo{Proof of Proposition 5.2} Denote
$$
F(\la)=\prod_{j=1}^k\prod_{\sigma=0}^{\th-1}
E_\th^*(\la;u^+_{j,\sigma})E_\th^*(\la;u^-_{j,\sigma}).
$$
By Proposition 5.1, for any $\la\in\Y$ we obtain
$$
\gather
F(\la)=\prod_{j=1}^k
E^*(\la_\th;-a_j-k\th-1)E^*(\la_\th;-a_j-k\th-\th)
\\
=\prod_{j=1}^k
E\bigl(\iota(\la_\th);-a_j-k\th-\tfrac 12\bigr)\,
E\bigl(\iota(\la_\th);-a_j-k\th-\th+\tfrac 12\bigr)
\\=
\prod_{j=1}^k
E\bigl((1-\xi)\iota(\la_\th);-x_j+O(1-\xi)\bigr)\,
E\bigl((1-\xi)\iota(\la_\th);-x_j+O(1-\xi)\bigr),
\endgather
$$
where in the last equality we used the 0-homogeneity of $E(\om;u)$.

We now split the average of $F(\la)$ with respect to $\wt M_{z,z',\th;\xi}$ into
two parts: over the Young diagrams $\la$ with
$(1-\xi)\cdot|\la|>C$ and $(1-\xi)\cdot|\la|\le C$ for some constant $C$.
The first one tends to zero as $C\to\infty$ uniformly in $\xi$ close to $1$.
Indeed, by Lemma 5.3(i),
$$
|F(\la)|\le e^{2\th k(1-\xi)|\la|/K}
$$
where we assume that $\min\{x_1,\dots,x_k\}>K$. By the hypothesis of the
proposition, we may choose $K$ as large as we need. Thus,
$$
\gather
\left|\sum_{n:\,(1-\xi)n>C }\,\sum_{|\la|=n} F(\la)\cdot  \wt
M_{z,z',\th;\xi}(\la)\right|\le\sum_{n:\,(1-\xi)n>C }\,\sup_{|\la|=n}
|F(\la)|\cdot  \wt M_{z,z',\th;\xi}(\Y_n)\\
\le (1-\xi)^t\sum_{n:\,(1-\xi)n>C } e^{2\th k(1-\xi) n/K} \frac{(t)_n}{n!}\xi^n.
\endgather
$$
For $\xi$ close to 1, $\xi^n=(1-(1-\xi))^n\le e^{-\const_1\cdot n(1-\xi)}$.
Further,
$$
(1-\xi)^t\,\frac{(t)_n}{n!}=(1-\xi)^t\,\frac{\Ga(t+n)}{\Ga(t)\Ga(n+1)}=
\frac{(1-\xi)^tn^{t-1}}{\Ga(t)}\,(1+O(n^{-1})).
$$
Hence, the first part of the average is bounded by
$$
\const_2\cdot (1-\xi)\sum_{n:(1-\xi)n>C}e^{\const_3\cdot n(1-\xi)}
$$
where $\const_3=2\th k/K-\const_1$. Choosing $K$ large enough, we make
$\const_3$ negative, and then the expression in question is bounded by
$$
\const_5\cdot \int_{C}^\infty e^{-\const_4\cdot s}ds, \qquad \const_4>0,
$$
which goes to 0 as $C\to\infty$.

The second part of the average has a limit as $\xi\nearrow 1$:
$$
\sum_{\la:\,(1-\xi)|\la|<C}F(\la)\wt M_{z,z',\th;\xi}(\la)\longrightarrow
\int\limits_{\om:\,\delta(\om)<C}\prod_{j=1}^k (E(\om_\th;-
x_j))^2\,\wt\M_{z,z',\th}(d\om).
$$
Indeed,
$$
F(\la)=\prod_{j=1}^k
E\bigl((1-\xi)\iota(\la_\th);-x_j+O(1-\xi)\bigr)\,
E\bigl((1-\xi)\iota(\la_\th);-x_j+O(1-\xi)\bigr)
$$
is uniformly close to
$$
\prod_{j=1}^k
(E\bigl(\om_\th;-x_j\bigr))^2,\qquad \om=(1-\xi)\iota(\la).
$$
by Lemma 5.3(ii). On the other hand, by Theorem 1.6 and Lemma 4.3, the image of
the measure $\wt M_{z,z',\th;\xi}$  under the map
$\la\mapsto\om=(1-\xi)\iota(\la)$, viewed as a measure on $\wOm$, weakly
converges to $\wt\M_{z,z',\th}$, as $\xi\nearrow1$.

Since
$$
\prod_{j=1}^k (E(\om_\th;-x_j))^2=\prod_{j=1}^k (E_\th(\om;-x_j))^{2\th},
$$
by Proposition 5.1, in order to conclude the proof of Proposition 5.2 it remains
to show that
$$
\int\limits_{\om:\,\delta(\om)>C}\prod_{j=1}^k (E(\om_\th;-
x_j))^2\,\wt\M_{z,z',\th}(d\om)
$$
converges to 0 as $C\to\infty$ uniformly in $\xi$ close to 1. This fact follows
from Lemma 5.3(i) similarly to the argument in the beginning of the proof. Note
that this estimate also justifies the convergence of the integral in the
right--hand side of \tht{5.1}. Another way to estimate the integral over
$\{\om:\delta(\om)>C\}$ is to directly use the integrability proved in Theorem
2.5.\qed
\enddemo

Recall that the lifted correlation functions $\wt\rho_k(x_1,\dots,x_k)$
(densities of the correlation measures $\wt\rho_k(dx)$) with positive arguments
$x_1,\dots,x_k$ were defined in \S1.

\proclaim{Theorem 5.5} For any $k=1,2,\dots$, and $x_1,\dots,x_k>0$,
$$
\gather
\wt\rho_k(x_1,\dots,x_k)=\prod_{j=1}^k \frac{\Ga(\th)}{\Ga(z-(j-1)\th)\Ga(z'-(j-
1)\th)}
\\ \times
(x_1\cdots x_k)^{z+z'+\th-1-2k\th}e^{-(x_1+\dots+x_k)}
\prod_{1\le i<j\le k}(x_i-x_j)^{2\th}
\\ \times
{}_2F_0^{(1/\th)}
\left(\frac{-z+k\th}\th\,,\frac{-z'+k\th}\th\,;\underbrace{-
\frac\th{x_1},\dots,-\frac{\th}{x_1}}_{2\th \text{  times
}},\,\dots,\underbrace{-\frac\th{x_k},\dots,-\frac\th{x_k}}_{2\th \text{  times
}}\right).
\endgather
$$
\endproclaim
\demo{Proof} The right--hand side is a real-analytic function in
$x_1,\dots,x_n>0$. Hence, by virtue of Proposition 1.9, it suffices to prove
the claim for $x_1,\dots,x_k\gg 0$.

On the other hand, for large $x_1,\dots,x_k$, the equality directly follows from
Proposition 4.2, Corollary 3.7, Proposition 5.2, and Theorem 2.5. Indeed,
Proposition 4.2 shows that the correlation measures $\wt \rho_k(dx)$ of
$\wt\M_{z,z',\th}$ are weakly approximated by their discrete counterparts ---
the correlation measures $\wt r_k^{(\xi)}$ of $M_{z,z',\th;\xi}$.
Further, Corollary 3.7 expresses the values of the discrete correlation measures
through averages of products of $E^*(\la;u)$.
Proposition 5.2 then shows that the weak limit of $\wt r_k^{(\xi)}$, if it
exists, must have the density equal to the integral
$$
\int\limits_{\om\in\wOm} \prod_{j=1}^k (E_\th(\om;-x_j))^{2\th}\, \wt\M_{z-
k\th,z'-k\th,\th}(d\om)
$$
(note the shift of $z,z'$ due to Corollary 3.7) times the limit of $(1-\xi)^{-k}
C'$ with $C'$ from Corollary 3.7 (the factor $(1-\xi)^{-k}$ comes from the
rescaling $\Z\to(1-\xi)\Z$). This limit is readily computed: for $a_i\sim
x_i/(1-\xi)$ as $\xi\nearrow 1$ we have
$$
\gather
\xi^{a_1+\dots+a_k+k(k+1)\th/2}\sim e^{-x_1-\dots-x_k},\\
\prod_{j=1}^k\frac
{\Ga(z+a_j+\th)\Ga(z'+a_j+\th)}{\Ga(a_j+k\th+1)\Ga(a_j+k\th+\th)}
\qquad\qquad\qquad\qquad\qquad\qquad\\
\qquad\qquad\qquad\qquad\qquad\qquad \sim
(1-\xi)^{-k(z+z'+\th-1)+2k^2\th}(x_1\cdots x_k)^{z+z'-2k\th+\th-1},\\
\prod_{1\le j<j'\le k}\prod_{\sigma=0}^{\th-1}((a_j-a_{j'})^2-\sigma^2)\sim
(1-\xi)^{k(k-1)\th}\prod_{1\le j<j'\le k}(x_j-x_{j'})^{2\th}.
\endgather
$$
Gathering these pieces together and using Theorem 2.5 we obtain the result.\qed
\enddemo

We can now invert the integral transform that relates the correlation functions
$\wt \rho_k$ of the lifted measure $\wt\M_{z,z',\th}$ and the correlation
functions $\rho_k$ of the initial measure $\M_{z,z',\th}$, see \S1.

It is convenient to introduce the notation, see \cite{GS}
$$
\frac {y_+^{c-1}}{\Ga(c)}=\cases \dfrac {y^{c-1}}{\Ga(c)}\,,& y>0,\\
0,&y\le 0.\endcases
$$
For $\Re c>0$ this is a locally integrable function. As a distribution, it
admits an analytic continuation in $c$ to the whole complex plane. In
particular, for $c=0$, $\frac {y_+^{c-1}}{\Ga(c)}$ is the delta-function at the
origin.

\proclaim{Theorem 5.6} For any $k=1,2,\dots$, and $x_1,\dots,x_k>0$
$$
\gather
\rho_k(x_1,\dots,x_k)=\Ga\left(\frac{zz'}{\th}\right)\cdot\prod_{j=1}^k
\frac{\Ga(\th)}{\Ga(z-(j-1)\th)\Ga(z'-(j-1)\th)}
\\ \times
(x_1\cdots x_k)^{z+z'+\th-1-2k\th}\,\frac{(1-|x|)_+^{c-1}}{\Ga(c)}
\prod_{1\le i<j\le k}(x_i-x_j)^{2\th}
\\ \times
{}_2\wh F_1^{(1/\th)}
\left(a,b;c;\underbrace{-\frac{\th(1-|x|)}{x_1},\dots,-\frac{\th(1-
|x|)}{x_1}}_{2\th \text{  times  }},\,\dots,\underbrace{-\frac{\th(1-
|x|)}{x_k},\dots,-\frac{\th(1-|x|)}{x_k}}_{2\th \text{  times  }}\right)
\endgather
$$
where $|x|=x_1+\dots+x_k$,
$$
a=\frac{-z+k\th}\th\,,\qquad
b=\frac{-z'+k\th}\th\,,\qquad c=ab\,\th.
$$
\endproclaim

Note that the expression above vanishes unless $|x|\le 1$. This agrees with the
fact that the correlation measure $\rho_k$ is supported by the set where $|x|\le
1$ as was mentioned in \S1.

\demo{Proof of Theorem 5.6} As was pointed out in \S1, the lifting \tht{1.4} is
invertible. Therefore, it suffices to check that \tht{1.4} holds with $\rho_k$
given by the formula above and $\wt \rho_k$ given by Theorem 5.5. We have
(recall that $t=zz'/\th$)
$$
\multline
\int_0^\infty\frac{s^{t-1}e^{-s}}{\Gamma(t)}\,
\rho_k\left(\frac{x_1}{s},\ldots,\frac{x_k}{s}\right)\frac {ds}{s^k}=
\prod_{j=1}^k \frac{\Ga(\th)}{\Ga(z-(j-1)\th)\Ga(z'-(j-1)\th)}\\
\times \int_0^{\infty}\left(\frac{x_1\cdots x_k}{s^k}\right)^{z+z'+\th-1-
2k\th}\cdot\frac{(s-|x|)_+^{c-1}}{s^{c-1}\Ga(c)}\cdot\frac{\prod_{1\le i<j\le
k}(x_i-x_j)^{2\th}}{s^{k(k-1)\th}}
\\ \times{}_2\wh F_1^{(1/\th)}
\left(a,b;c;\underbrace{-\frac{\th(s-|x|)}{x_1},\dots,-\frac{\th(s-
|x|)}{x_1}}_{2\th \text{  times  }},\,\dots,\underbrace{-\frac{\th(s-
|x|)}{x_k},\dots,-\frac{\th(s-|x|)}{x_k}}_{2\th \text{  times  }}\right)\\
\times s^{t-1-k}e^{-s}ds.
\endmultline
$$
Making the change of variable $s-|x|\to s$ and using \tht{2.5}, we obtain the
result.\qed
\enddemo

\example{Remark 5.7} Assume, as in Remarks 1.10 and 2.6, that $z=m\th$,
$m=1,2,\dots$, and $z'>(m-1)\th$. Then Theorems 5.5 and 5.6 show that $\rho_k$
and $\wt\rho_k$ vanish identically for $k\ge m+1$, which agrees with the fact
that the measures $\M_{z,z',\th}$ and $\wt\M_{z,z',\th}$ live on the subsets of
$\Om$ and $\wOm$ with no more than $m$ nonzero alpha-coordinates. (The
vanishing is caused by the gamma--prefactors.)

The $m$th correlation function gives the distribution function for
$\al_1,\dots,\al_m$ given by \tht{1.5} and \tht{1.6}. Further, the formulas of
Theorems 5.6 and 5.5 with  $k<m$ provide the correlation functions for the
$m$-particle Laguerre ensemble \tht{1.6} and its simplex analog \tht{1.5}.
\endexample

\example{Remark 5.8} Theorems 5.5, 5.6, and Remark 2.7
provide integral representations for the density functions $\widetilde\rho_1$
and $\rho_1$ which involve only elementary functions. A similar integral
representations has been used in \cite{BF, \S5.3} for (saddle point)
asymptotic analysis of the density function in the Hermite ensemble when the
number of particles goes to infinity.
\endexample

\head 6. Asymptotics of the correlation functions at the origin
\endhead

In this section we compute the asymptotics of the correlation functions
$\rho_k(x)$ and $\wt\rho_k(x)$ when $x_1,\dots,x_k\to +0$. In the variables
$y_i=-\ln x_i$ the answer is translation invariant and is the same for both
lifted and non-lifted correlation functions. This limit transition is similar to
the bulk scaling limit in random matrix models.

We will need certain multivariate special functions
$\varphi_s^{(\nu)}(x_1,\dots,x_l)$, $s\in\C^l$, $x\in(\Bbb R_{>0})^l$. These
functions are symmetric
with respect to permutations of $\{x_i\}$ and  generalize the normalized Jack
polynomials $P_\la^{(\nu)}(x_1,\dots,x_l)/P_\la^{(\nu)}(1,\dots,1)$: if
$s=\la+\rho$, where
$$
\rho=\nu\left(\frac{l-1}2,\frac{l-3}2,\dots,-\frac{l-3}2,-\frac{l-1}2\right),
$$
then these two functions coincide.

The functions $\varphi_s^{(\nu)}$ can be defined as symmetric, normalized at
$(1,\dots,1)$ eigenfunctions of the Sekiguchi system of differential operators
with appropriate eigenvalues depending on $s$, see \cite{Sek} and also
\cite{Ma2}. The functions $\varphi_s^{(\nu)}$ are symmetric with respect
to the permutations of $\{s_i\}$.

When $\nu=1/2,1,2$, the functions $\varphi_s^{(\nu)}$ are spherical functions
for the symmetric space
$GL(l,\Bbb F)/U(l,\Bbb F)$, where $\Bbb F=\R,\C,\Bbb H$, respectively, and they
admit a matrix integral representation, see \cite{FK, chapter XIV,
\S3}.

In the case $\theta=1$ the spherical function is given by the explicit formula
$$
\varphi_{s_1,\dots,s_l}^{(1)}(x_1,\dots,x_l)=
0!1!\cdots (l-1)!\cdot \frac{(x_1\dots x_l)^{\frac{l-
1}2}\det[x_i^{s_j}]}{\prod_{i<j}(x_i-x_j)(s_i-s_j)}\,.
$$

\proclaim{Theorem 6.1} For any $k=1,2,\dots$, the image of the correlation
measure $\rho_k(dx)$ or $\wt\rho_k(dx)$ under the change of variables
$$
x_i=e^{-y_i-T}, \qquad i=1,\dots,k,
$$
converges, as $T\to+\infty$, to
$$
C\cdot \prod_{1\le i<j\le k}(e^{-y_i}-e^{-y_j})^{2\th}\cdot\varphi_s^{(1/\th)}
(\,\underbrace{e^{-y_1},\dots,e^{-y_1}}_{2\th\text{
times}},\dots,\underbrace{e^{-y_k},\dots,e^{-y_k}}_{2\th\text{  times}})\,dy,
\tag 6.1
$$
where
$$
\gather
C=\prod_{j=0}^{k-1}\frac{\Ga(j\th+1)\Ga(\th)\Ga(j\th+z-z'+1)\Ga(j\th+z'-
z+1)}{\Ga(j\th+k+1)\Ga(j\th-z+1)\Ga(j\th-z'+1)\Ga(z-j\th)\Ga(z'-j\th)}\,,
\\
s=(s'_1,\dots,s'_{k\th},s_1'',\dots,s_{k\th}''),\\
s_j'=\frac{z'-z-2j+\th+1}{2\th}\,,\quad s_j''=\frac{z-z'-
2j+\th+1}{2\th}\,,\qquad
j=1,\dots,k\th.
\endgather
$$
\endproclaim

Note that the measure \tht{6.1} is translation invariant. Indeed, this follows
from the fact that
$$
\varphi^{(\nu)}_{s}(a\cdot x_1,\dots,a\cdot
x_l)=a^{|s|}\varphi^{(\nu)}_{s}(x_1,\dots,x_l), \qquad |s|=s_1+\dots+s_l,
$$
for any $a>0$ and $l=1,2,\dots$\,.

The result for $\th=1$ was proved in \cite{P.III}. A stronger result
involving joint correlation functions of $\{\al_i\}$ and $\{\be_i\}$ (also for
$\th=1$) was proved in \cite{P.V}.

The proof of Theorem 6.1 is based on multivariate Mellin-Barnes integral
representations of ${}_2F_0^{(\nu)}$ and ${}_2\wh F_1^{(\nu)}$. The details will
appear elsewhere.

\vskip 2 true cm

{\smc A.~Borodin}: Mathematics 253-37, Caltech, Pasadena, CA 91125, U.S.A.

E-mail address: {\tt borodin\@caltech.edu}

\medskip

{\smc G.~Olshanski}: Dobrushin Mathematics Laboratory, Institute for
Information Transmission Problems, Bolshoy Karetny 19, 101447 Moscow GSP-4,
RUSSIA.

E-mail address: {\tt olsh\@online.ru}

\enddocument